\documentclass[twocolumn]{aastex6}

\usepackage{amsmath}

\defcitealias{bailin09}{BH09}

\newcommand{\MGMC}{{\ensuremath{M_{\mathrm{cloud}}}}}
\newcommand{\MGC}{{\ensuremath{M_{\mathrm{GC}}}}}
\newcommand{\Nclump}{{\ensuremath{N_{\mathrm{clump}}}}}
\newcommand{\Mclump}{{\ensuremath{M_{\mathrm{clump}}}}}
\newcommand{\tform}{{\ensuremath{t_{\mathrm{form}}}}}
\newcommand{\Msun}{{\ensuremath{M_{\sun}}}}

\begin{document}

\shorttitle{Clumpy Self-Enrichment in Globular Clusters}
\shortauthors{Bailin}

\title{A Model for Clumpy Self-Enrichment in Globular Clusters}
\author{Jeremy Bailin}
\affil{Department of Physics and Astronomy, University of Alabama, Box 870324, Tuscaloosa, AL, 35487-0324 USA}
\and
\affil{Steward Observatory, University of Arizona, 933 N Cherry Ave., Tucson, AZ,  85721-0065 USA}
\email{jbailin@ua.edu}

\begin{abstract}
Detailed observations of globular clusters (GCs) have revealed evidence of self-enrichment: some of the heavy elements that we see in stars today were produced by cluster stars themselves. Moreover, GCs have internal subpopulations with different elemental abundances, including, in some cases, in elements such as iron that are produced by supernovae. This paper presents a theoretical model for GC formation motivated by observations of Milky Way star forming regions and simulations of star formation, where giant molecular clouds fragment into multiple clumps which undergo star formation at slightly different times. Core collapse supernovae from earlier-forming clumps can enrich later-forming clumps to the degree that the ejecta can be retained within the gravitational potential well, resulting in subpopulations with different total metallicities once the clumps merge to form the final cluster. The model matches the mass-metallicity relation seen in GC populations around massive elliptical galaxies, and predicts metallicity spreads within clusters in excellent agreement with those seen in Milky Way GCs, even for those whose internal abundance spreads are so large that their entire identity as a GC is in question. The internal metallicity spread serves as an excellent measurement of how much self-enrichment has occurred in a cluster, a result that is very robust to variation in the model parameters.
\end{abstract}

\section{Introduction}
Until relatively recently, globular clusters (GCs) had been considered to be, to good approximation, simple stellar populations (SSPs) -- populations of stars with identical ages and chemical abundances, due to a simultaneous formation from a well-mixed parent giant molecular cloud (GMC).

However, as observations have improved, it has become increasingly apparent that this picture is inaccurate, or at least incomplete. In particular, there is now significant evidence that the composition of some GC stars retain traces of the effects of other stars within the cluster. In other words, not all of the heavy elements come from \textit{pre-enrichment} of the protocluster cloud; some come from \textit{self-enrichment} by the cluster itself. There are two distinct lines of evidence for this picture: the presence of multiple stellar populations, and the ``blue tilt''.

Virtually all GCs with sufficiently accurate photometric and/or spectroscopic data show two or more populations, which manifest as multiple main sequences, main sequence turnoffs, subgiant branches, and/or red giant branches (e.g. \citealp{piotto15}; see \citealp{bastian18} for a recent review).
These different populations are always associated with different abundances of light elements such as C, O, N, and Na, sometimes with different abundances of He, and in a few clusters spreads in Fe. The abundance patterns \citep[e.g., the ubiquitious O-Na anti-correlation;][]{carretta09} are impossible to generate from inhomogeneities in the protocluster cloud, and must be due to enrichment during GC formation.

The second line of evidence is the ``blue tilt'', a mass-metallicity relation (MMR) that is observed for the most massive metal-poor GCs around most (but not all) massive galaxies with large numbers of GCs \citep[e.g.][]{harris06,mieske06}. In a pure pre-enrichment model, there is no causal mechanism whereby the past chemical enrichment history of the protocluster cloud, i.e. the metallicity, can know how large a GC will eventually form in it. Self-enrichment solves this problem: more massive clouds with deeper gravitational potential wells can retain supernova ejecta, which can be incorporated into stars that have not yet finished forming, increasing the metallicity of GC stars. Therefore, the stars of the cluster have a direct influence on the cluster metallicity.

The mass of the cluster is also relevant for multiple populations. Although even the smallest GCs have multiple populations traced by their light elements, \citet{milone17} showed that the photometric width of the red giant branch, which is a measure of total metallicity spread, correlates strongly with GC mass at a given total metallicity. This suggests that multiple populations and the MMR may have a common self-enrichment origin.

The \citeauthor{bailin09} (\citeyear{bailin09}; hereafter \citetalias{bailin09}) self-enrichment model was inspired by the MMR. Rather than all stars forming at once, as in an SSP, in \citetalias{bailin09} all stars \textbf{begin} to form at once. However, the dramatically shorter formation and evolution timescales of massive stars that explode as core collapse supernovae (CCSNe) compared to the formation timescales of low-mass stars that survive 10~Gyr later, allow the high mass stars to explode and pollute the low mass stars while they are still forming.

The \citetalias{bailin09} model provides a good qualitative picture of the MMR in GCs. The quantitative agreement is fair, although the predicted MMR kicks in at higher mass than is observed; the quantitative agreement becomes excellent with plausible modifications to the model free parameters \citep{mieske10} or aggressive dynamical mass loss \citep{goudfrooij14}.
However, while this model is a significant improvement over an SSP, it still treats each GC as internally homogeneous. It is therefore unable to shed light on multiple populations, which are one of the main tracers of the self-enrichment process, and is poorly inspired by the actual structure of massive star forming regions, which both observations and simulations tell us is complex, clumpy, and non-simultaneous.

While true local analogs to the extremely high-density high-intensity star formation regions that must have given birth to GCs do not exist, the Carina Nebula complex is a region with large amounts of high-mass star formation that is close enough to us that its structure can be divined, and is therefore a good proxy. As described in detail by \citet{smithbrooks08}, Carina is a giant star forming region with a number of spatially-distinct subclusters (Tr 14, 15, 16; Bo 10, 11) with ages that range from $\sim 3$~Myr stars that are about to explode ($\eta$~Car) to regions that are still collapsing to form new stars today and into the future. It therefore provides direct evidence that star formation is clumpy, extended in time, and that CCSNe can impact stars formed within the same giant molecular cloud (GMC).

Theoretical hydrodynamic simulations of molecular cloud-scale star formation predict such clumpiness. In these simulations, gravitationally bound substructures form from the initial turbulence of the protocluster cloud. Star formation begins within these substructures at different times, before the subclusters eventually merge \citep{bate09}. The final cluster is the conglomeration of a number of pieces that were originally subclumps within the natal cloud.

In this paper, I extend the \citetalias{bailin09} model to the case where the protocluster cloud fragments into a number of semi-independent clumps, which can both locally self-enrich and enrich each other, before merging together to form the final cluster, as inspired by both observations and simulations of star formation. 
This paper focuses on the total metal abundance in GCs due to CCSNe, as traced by iron, but the model is intentionally constructed so that it can be easily modified in the future to follow the enrichment of specific elements formed in different stages of stellar lives.
Iron provides the ideal first testbed for the viability of the self-enrichment model because, on the relevant timescales, CCSNe are the only important nucleosynthetic source.
In contrast, the lighter elements that show evidence for self enrichment (e.g., C, O, N, Na) can also be produced by asymptotic giant branch stars, fast rotating massive stars, binaries, and/or very massive stars. Therefore, predictions for those elements are much more sensitive to uncertain yields from poorly-constrained stellar processes than the (comparatively!) better understood CCSNe. Once the viability of this model is demonstrated, future work will extend the model to these additional enrichment mechanisms in order to understand the multiple populations seen in light element abundances.

In Section~\ref{sec:model-details}, I present the details of the model, followed by a discussion of the constraints on the free parameters of the model and the choices for the fiducial parameter values (Section~\ref{sec:fiducial-parameters}). The results are presented in Section~\ref{sec:results}, including an exploration of how the results depend on the free parameters. Discussion and conclusions are presented in Section~\ref{sec:conclusions}.
The code used to generate these models is released in \citet{bailin18-gczcse-software}.

\section{Model Details}\label{sec:model-details}

\begin{figure*}
\plotone{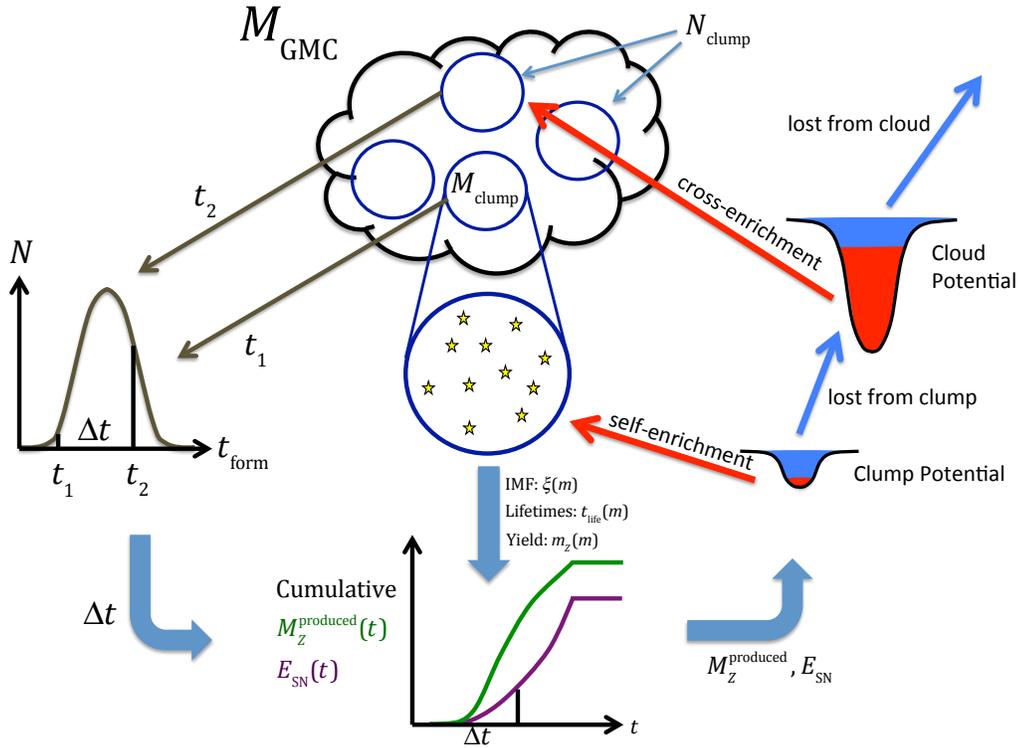}
\caption{\label{fig:schematic}%
Schematic view of the clumpy self-enrichment model. The protocluster cloud fragments
into a number of clumps, which begin to form stars at a range of times drawn from
a normal distribution. After an amount of time $\Delta t$, the stellar population
has produced a total amount of metals, Cumulative $M^{\mathrm{produced}}_Z(t)$,
and supernova energy, $E_{\mathrm{SN}}(t)$, due to the supernovae that have gone off.
The form of these functions depend on the initial mass function (IMF) $\xi(m)$, the stellar mass-lifetime
relation $t_{\mathrm{life}}(m)$, and the supernova metal yield $m_Z(m)$.
The metals get mixed within the clump, and a fraction is unbound
depending on the ratio of $E_{\mathrm{SN}}$ to the depth of the clump potential well;
the remainder self-enriches the less-massive still-forming stars within the clump
itself. Metals and energy that overflow the clump become mixed with the cloud
as a whole, and the fraction that do not overflow the deeper cloud potential
well is available to cross-enrich clumps that have not yet begun forming stars.}
\end{figure*}

Figure~\ref{fig:schematic} shows a schematic overview of the clumpy self-enrichment model.
In this model, the cloud fragments into distinct clumps, which each form an
SSP at a distinct time. Each clump undergoes self-enrichment due to its own
massive stars, but can also contaminate later-forming clumps with metals that
overflow the clump's potential well within the cloud (``cross-enrichment''). The amount of metals
available for cross-enrichment, and the amount of energy that can push those
metals out of the clump or even out of the entire cloud, build up with time as
more stars explode. The final metal distribution of stars within the GC is the
mass-weighted combination of the metallicities of the SSPs formed in each clump.

In detail, we begin with a protocluster cloud of mass \MGMC. When
star formation occurs, the cloud fragments into \Nclump\ distinct clumps.
The number of clumps scales with the cloud mass as
\begin{equation}
	\Nclump = \Nclump_{\mathrm{,ref}} \left(\frac{\MGMC}{\MGMC_{\mathrm{,ref}}}\right)^s
	\label{eq:Nclump scaling}
\end{equation}
such that $s=0$ means that each cloud fragments into the same number of clumps,
while $s=1$ means that each clump has the same mass, normalized
at $\MGMC_{\mathrm{,ref}} \equiv 10^5~\Msun$. The number of clumps 
for a given cloud is drawn from a Poisson distribution
whose mean is given by equation~(\ref{eq:Nclump scaling}),
with the caveat that $\Nclump \ge 1$ (i.e. if the random number generator draws $0$, $\Nclump$ is set to $1$).
The total fraction of the cloud that is contained in star-forming clumps (``dense gas'')
is $f_{\mathrm{DG}}$; the remainder is in diffuse gas.

The mean mass per clump is
\begin{equation}
	\left<\Mclump\right> = \frac{f_{\mathrm{DG}} \MGMC}{\Nclump}.
\end{equation}
The mass of each clump is drawn from a lognormal distribution of logarithmic width
$\sigma_M$ and mean $\left<\Mclump\right>$ \citep{ellsworthbowers15}.
Because of random sampling, it is possible for the sum of the clump masses to be larger
than the original cloud mass; to prevent this, after the clump masses have been randomly
drawn, all $\Mclump_{,i}$ are retroactively rescaled by a constant factor to ensure that
\begin{equation}
    \sum_i \Mclump_{,i} = f_{\mathrm{DG}} \MGMC.
\end{equation}
Each clump is assigned a formation time \tform\ drawn from a normal distribution of
width $\sigma_{\mathrm{form}}$.

Star formation proceeds with a global efficiency $f_*$,
so the initial stellar mass of the globular cluster is $\MGC = f_* \MGMC$.
Stars form within each clump according to a power-law IMF:
\begin{equation}
    \frac{dn}{dm} \equiv \xi(m) = A m^{\alpha}
\end{equation}
within the mass range $M_{\mathrm{min}} < m < M_{\mathrm{max}}$.
Although the low-mass IMF is known to turn over \citep[e.g.][]{chabrier03}, the correct number of CCSNe
as a function of progenitor mass only requires the correct high-mass slope and an appropriate
low mass cutoff $M_{\mathrm{min}}$ to normalize the fraction of mass within the population
contained in stars with masses $m>9~\Msun$. As in \citetalias{bailin09}, I adopt
$M_{\mathrm{min}}=0.30~\Msun$ and $M_{\mathrm{max}}=100~\Msun$.

Clumps are labeled in order of increasing \tform. Each clump is assumed to form
long-lived stars with a uniform metallicity. For clump $i$, there are three
sources of metals:
\begin{equation}\label{eq:MZ sum}
    M_{Z,i} = M^{\mathrm{pre}}_{Z} + M^{\mathrm{self}}_{Z,i} + M^{\mathrm{cross}}_{Z, j<i}
\end{equation}
where $M^{\mathrm{pre}}_Z$ is the pre-enrichment of the
protocluster cloud before fragmentation occurs, $M^{\mathrm{self}}_Z$ comes from
self-enrichment of the low-mass stars in the clump by its own high-mass stars,
and $M^{\mathrm{cross}}_{Z, j<i}$ is the total enrichment due to metals produced
from clumps $j$ that formed before clump $i$, which is spread throughout the cloud.

Self-enrichment within the clump proceeds mostly as in \citetalias{bailin09},
but applied to just the clump instead of the whole cloud. Briefly, the total mass
of metals $M^{\mathrm{produced}}_{Z,\mathrm{tot}}$, total ejecta mass $M^{\mathrm{produced}}_{\mathrm{tot}}$, and supernova energy $E_{\mathrm{SN,tot}}$
produced by the CCSNe from all high mass stars in the clump
are available to mix within the clump. 
The density profile of the clump is a singular isothermal
sphere truncated at $r=r_t$:
\begin{equation}
    \rho(r) = \left\{ \begin{array}{l@{\quad}l}
        \frac{\Mclump}{4 \pi r_t^3} \left(\frac{r}{r_t}\right)^{-2} & r \le r_t \\
        0 & \textrm{otherwise} \\ \end{array} \right.
\end{equation}
for which the potential is
\begin{equation}
    \Phi(r) = \left\{ \begin{array}{l@{\quad}l}
        \frac{G \Mclump}{r_t} \ln\left(\frac{r}{r_t}\right) & r \le r_t \\
        0 & \textrm{otherwise} \\ \end{array} \right.
\end{equation}
Larson's relations \citep{larson81}, based on observed GMCs, imply that clump surface densities are approximately
constant, so the clump radius scales as
\begin{equation}
    r_t = \sqrt{\frac{\Mclump}{\pi \Sigma_{\mathrm{clump}}}}.
\end{equation}
Then the fraction of material that can be retained  is equal to the fraction of the cloud that
lies deep enough within the potential that $E_{\mathrm{SN,tot}}/\Mclump < \Phi(r)$,
and is equal to
\begin{equation}\label{eq:fZ SE}
    f_{\mathrm{retained}} = \exp\left(-\frac{E_{\mathrm{SN,tot}} r_t}{G \Mclump^2}\right)
\end{equation}
while the fraction of material ejected is
\begin{equation}
    f_{\mathrm{ejected}} = 1 - f_{\mathrm{retained}}
\end{equation}

The CCSN ejecta may mix extensively with the existing gas in the cloud (as in \citetalias{bailin09}), or most of the ejecta could escape directly without mixing.
I parameterize the efficiency
of mixing as $f_{\mathrm{mix}} \in [0,1]$ which linearly interpolates the mass of ejected metals $M^{\mathrm{ejected}}_Z$ between two extremes:
\begin{description}
 \item[Minimal mixing] If $f_{\mathrm{mix}}=0$, then the mass that escapes from the clump consists entirely of CCSN ejecta, with the limit that clump cannot release more metals than were produced.
 \item[Maximal mixing] If $f_{\mathrm{mix}}=1$, then the CCSN ejecta are assumed to fully mix within the clump, so the fraction of produced metals that escape the clump is equal to the fraction of the total clump material that escapes.
 \end{description}
Formally,
\begin{multline}\label{eq:mixing}
  M^{\mathrm{ejected}}_Z = f_{\mathrm{mix}} f_{\mathrm{ejected}}\, M^{\mathrm{produced}}_{Z,\mathrm{tot}} + \\
  \left(1 - f_{\mathrm{mix}}\right) \min\left( f_{\mathrm{ejected}} \Mclump Z^{\mathrm{produced}}_{\mathrm{tot}}, M^{\mathrm{produced}}_{Z,\mathrm{tot}} \right)
\end{multline}
and the total ejected mass is
\begin{equation}
  M^{\mathrm{ejected}} = f_{\mathrm{ejected}} \Mclump
\end{equation}
where
\begin{equation}\label{eq:Zproduced SE}
  Z^{\mathrm{produced}}_{\mathrm{tot}} = \frac{M^{\mathrm{produced}}_{Z,\mathrm{tot}}}{M^{\mathrm{produced}}_{\mathrm{tot}}}
\end{equation}

The self-enriched metal mass is
\begin{equation}\label{eq:Zself}
    M^{\mathrm{self}}_Z = M^{\mathrm{produced}}_{Z,\mathrm{tot}} - M^{\mathrm{ejected}}_Z
\end{equation}
while the material that overflows the clump potential is available to cross-enrich
later-forming clumps within the cloud. The total amounts of metal, mass, and energy
contributed to the cloud after time $t$, $M_{Z,j\rightarrow\mathrm{ejected}}(t)$, $M_{j\rightarrow\mathrm{ejected}}(t)$, and $E_{\mathrm{SN},j\rightarrow\mathrm{ejected}}(t)$ respectively, are calculated as in equations~(\ref{eq:fZ SE}) through (\ref{eq:Zproduced SE}), but replacing $M^{\mathrm{produced}}_{Z,\mathrm{tot}}$, $M^{\mathrm{produced}}_{\mathrm{tot}}$, and $E_{\mathrm{SN,tot}}$ with
\begin{equation}
  M^{\mathrm{produced}}_{Z}(t) = \sum_{m > m_{\mathrm{max}}(t)} m_Z(m),
\end{equation}
\begin{equation}
  M^{\mathrm{produced}}(t) = \sum_{m > m_{\mathrm{max}}(t)} m_{\mathrm{ejecta}}(m),
\end{equation}
and
\begin{equation}
    E_{\mathrm{SN}}(t) = \sum_{m > m_{\mathrm{max}}(t)} E_{\mathrm{SN}}
\end{equation}
respectively, where
the most massive star still alive after time $t$ that explodes as a CCSN is $m_{\mathrm{max}}(t)$,
$m_Z(m)$ is the metal mass produced by a supernova of a star of initial mass $m$,
$m_{\mathrm{ejecta}}(m) \equiv m - m_{\mathrm{remnant}}(m)$ is the total mass of supernova ejecta from a star of mass $m$,
and $E_{\mathrm{SN}}$ is the energy produced by a single CCSN, assumed to be $10^{51}$~erg.

The adopted stellar mass-lifetime relation $m_{\mathrm{max}}(t)$ is a polynomial fit to 
MIST models \citep{choi16}; see Appendix~\ref{sec:stellar-lifetimes}. Different CCSN yield
functions $m_Z(m)$ and $m_{\mathrm{ejecta}}(m)$ can be adopted; see Appendix~\ref{sec:yields}.

If the clumps contribute sufficient energy, some of the metals can be lost from the cloud completely, and are therefore not available for cross-enrichment.
When considering cloud $i$, the relevant metal masses, total masses, and energies are
\begin{equation}
  M^{\mathrm{produced}}_{Z,j<i} = \sum_{j<i} M_{Z, j\rightarrow \mathrm{ejected}}(\Delta t_{ji}),
\end{equation}
\begin{equation}
  M^{\mathrm{produced}}_{j<i} = \sum_{j<i} M_{ j\rightarrow \mathrm{ejected}}(\Delta t_{ji}),
\end{equation}
and
\begin{equation}
    E_{\mathrm{SN},j<i} = \sum_{j<i} E_{\mathrm{SN},j \rightarrow \mathrm{ejected}}(\Delta t_{ji}).
\end{equation}
where clump $j$ formed $\Delta t_{ji}$ before clump $i$.
In analogy with equations~(\ref{eq:fZ SE}) through (\ref{eq:Zself}), the metal mass available within the cloud
for cross-enriching clump $i$ is then calculated as
\begin{equation}
    f_{\mathrm{retained},j<i} = \exp\left(-\frac{E_{\mathrm{SN},j<i} r_t}{G \MGMC^2}\right)
\end{equation}
\begin{equation}
    f_{\mathrm{ejected},j<i} = 1 - f_{\mathrm{retained},j<i}
\end{equation}
\begin{multline}\label{eq:mixingcloud}
  M^{\mathrm{ejected}}_{Z,j<i} = f_{\mathrm{mix}} f_{\mathrm{ejected},j<i}\, M^{\mathrm{produced}}_{Z,j<i}
    + \\
    \left(1 - f_{\mathrm{mix}}\right) \min\left( f_{\mathrm{ejected},j<i} \MGMC Z^{\mathrm{produced}}_{j<i}, M^{\mathrm{produced}}_{Z,j<i} \right)
\end{multline}
where the total metallicity of the material that has been lost from the clumps,
\begin{equation}
  Z^{\mathrm{produced}}_{j<i} = \frac{M^{\mathrm{produced}}_{Z,j<i}}{M^{\mathrm{produced}}_{j<i}},
\end{equation}
can be significantly lower than the metallicity of pure CCSN ejecta $Z^{\mathrm{produced}}_{\mathrm{tot}}$ because substantial amounts of unenriched clump material is part of the ejected mass.

Finally, the mass of metals that remain are split evenly throughout the cloud, so the extra metals contributed to clump $i$ via cross-enrichment is
\begin{equation}
  M^{\mathrm{cross}}_{Z,j<i} = \frac{M^{\mathrm{produced}}_{Z,j<i} - M^{\mathrm{ejected}}_{Z,j<i}}{\Nclump}
\end{equation}
The metallicity of the long-lived stellar population formed in the clump is
\begin{equation}
    Z_i = \frac{M_{Z,i}}{\Mclump_{,i}}
\end{equation}
where $M_{Z,i}$ is the sum of all contributions (equation~\ref{eq:MZ sum}).

Total GC metallicity is usually traced by [Fe/H]. For typical GC values of [$\alpha$/Fe]$\approx 0.3$
\citep[e.g.][]{kirby08}, these are related by
\begin{equation}
    \mathrm{[Fe/H]} = \log Z / Z_{\sun} - 0.25
\end{equation}
where $Z_{\sun}=0.016$ \citepalias{bailin09}.
In the future this model
will be extended to follow elements individually, at which point this assumption will no longer
be necessary.

This model has been instantiated as a Python code, \texttt{GCZCSE} (Globular Cluster metallicity (Z) Clumpy Self-Enrichment)
that is released in \citet{bailin18-gczcse-software}. GCZCSE produces a Monte Carlo sampling of clusters over a range of masses,
and uses the \texttt{multiprocessing} module for single-node parallelization.

\section{Fiducial Model Parameters}\label{sec:fiducial-parameters}

\begin{table}
\caption{Fiducial Parameter Values\label{table:fiducial parms}}
\begin{tabular}{lc}
Parameter & Value \\ \hline
$f_*$ & $0.3$ \\
$\alpha$ & $-2.35$ \\
$f_{\mathrm{DG}}$ & $1.0$ \\
$s$ & $0.60$ \\
$\Nclump_{,\mathrm{ref}}$ & $3.4$ \\
$\sigma_M$ & $0.1$ \\
$\Sigma_{\mathrm{clump}}$ & $100~\Msun~\mathrm{pc}^{-2}$\\
$\sigma_{\mathrm{tform}}$ & $10 \times 10^6$~yr \\
$f_{\mathrm{mix}}$ & $0.2$ \\
CCSN Yields & \citet{nomoto13} \\
\end{tabular}
\end{table}

The free parameters of the model are listed in Table~\ref{table:fiducial parms}, along with their
values for the fiducial model. There are two parameters related to the shape and normalization
of the IMF, four parameters related to fragmentation, which determine
the masses of the clumps, one that is related to the mass-radius relation of the clumps, one that
is related to the duration of star formation, one related to mixing, and one related to CCSN yields. Justification for this choice of parameters is given below.

Any particular run of \texttt{GCZCSE} also requires a pre-enrichment level for the protocluster gas cloud.

\subsection{Initial Mass Function}
All CCSNe come from the high-mass end of the IMF, which is usually considered to be independent of environment and to have a power-law slope around the \citet{salpeter55} value of $\alpha=-2.35$ \citep{chabrier03}.
Recent observations hint that particularly vigorous star formation regions could have more top-heavy IMFs; for example, \citet{motte18} find that the core mass function (which may translate into the IMF when shifted by a factor $f_*$; \citealp{alves07}) in a Galactic mini-starburst may have a slope of $\sim -1.9$,
while \citet{zhang18} argue that dusty sturbursts have IMF slopes shallower than $-1.7$.
For the fiducial model, I adopt the ``universal'' Salpeter IMF, but explore the impact of IMF variations in Section~\ref{sec:parameter-dependence}.

The efficiency of star formation can be estimated from
Galactic embedded clusters, which evolve towards an apparently universal star formation efficiency of $f_* = 0.3$ \citep{lada03,marks08}, which is adopted as the fiducial value.

\subsection{Clumping Parameters}
The fraction of the cloud that ends up in clumps, often referred to as the fraction
of dense gas $f_{\mathrm{DG}}$, depends observationally
on the efficiency of star formation \citep{lada12,roccatagliata13}, with the highest
efficiency star formation corresponding to $f_{\mathrm{DG}} \approx 1$. In order
to form such high stellar densities, GCs must have among the highest star formation
efficiencies of any star formation events, and so I adopt $f_{\mathrm{DG}}=1.0$ as the
fiducial value.

\begin{figure}
\plotone{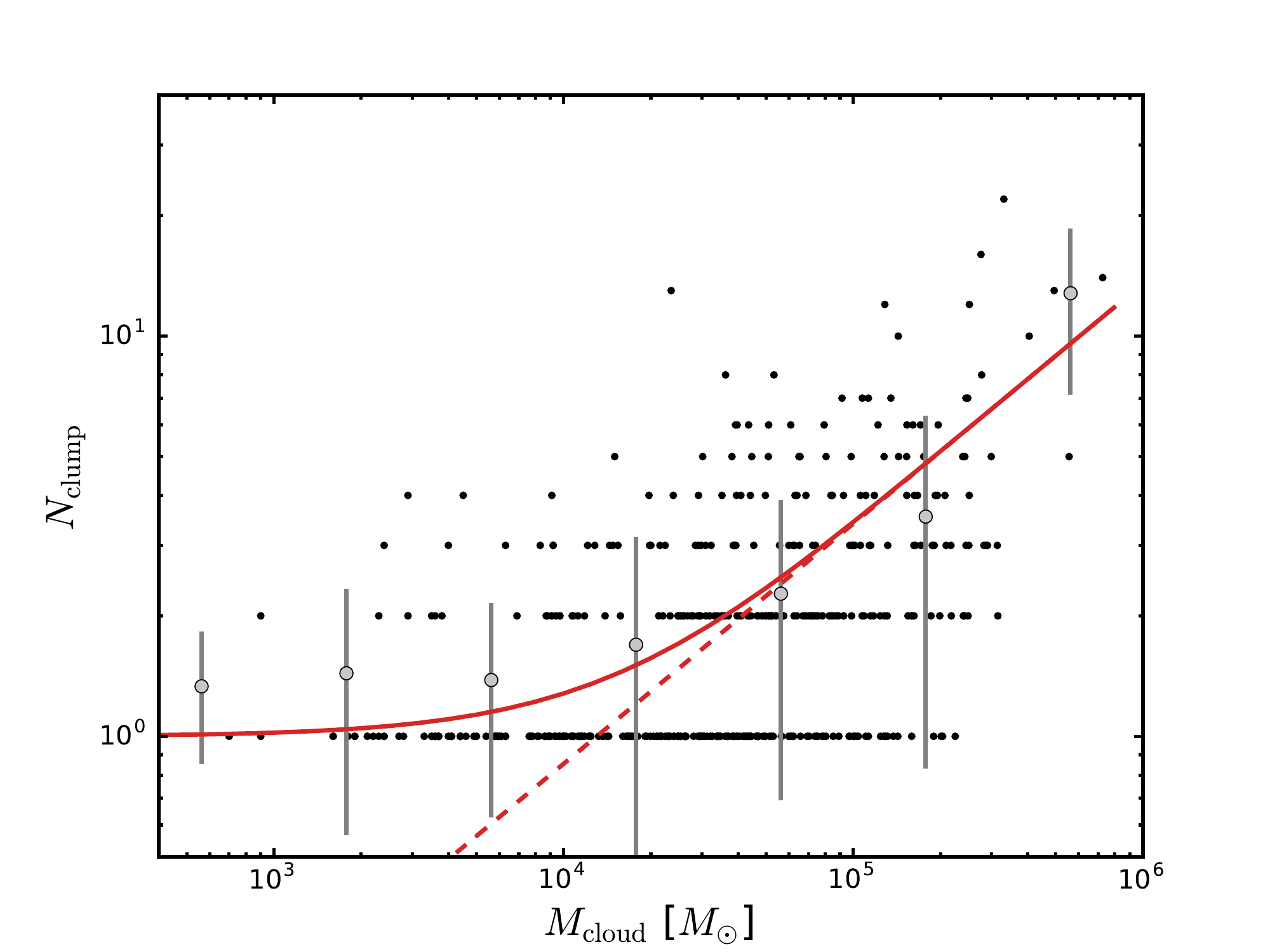}
\caption{\label{fig:fragmentation scaling}%
Number of BGPS clumps per GMC. Data points are from \citet{battisti14}. Gray error bars
denote the mean and standard deviation in bins of width $0.5$~dex. The adopted scaling
is shown in red; the dashed line is the power law given in equation~(\ref{eq:Nclump scaling}),
while the solid line is the fit that takes into
account the fact that $\Nclump \ge 1$.}
\end{figure}

The number of clumps into which the cloud fragments, and its dependence on \MGMC, can be
derived from the work of \citet{battisti14}, who catalogued GMCs using $^{13}\mathrm{CO}~J=1-0$
emission and clumps via 1.1~mm dust continuum emission in the Bolocam Galactic
Plane Survey (BGPS). The data are plotted in Figure~\ref{fig:fragmentation scaling}, along
with the mean and standard deviation within bins of width $0.5$~dex. A modified power-law
was fitted to the binned data, and is shown as the red line. The modification accounts for the
fact that $\Nclump \ge 1$; if the value of \Nclump\ that is derived from the power-law
(equation~\ref{eq:Nclump scaling}, shown as the red dashed line) is $\nu$,
then the Poisson distribution has a probability
$P_{\nu}(0) = e^{-\nu}$ of being equal to $0$. As \Nclump\ is set to 1 in those cases, the actual
mean becomes $\nu + e^{-\nu}$. The best-fit value of $s=0.60$ implies
that massive clouds both contain more clumps than small clouds ($s>0$),
and each clump is itself more massive ($s<1$).
The normalization, i.e. the mean number of clumps in a
$10^5~\Msun$ cloud, is $\Nclump_{\mathrm{,ref}}=3.4$.

The width of the distribution of clump masses within a cloud is fairly uncertain.
\citet{ellsworthbowers15} used the BGPS to derive the total clump mass function,
and found a wide distribution, with lognormal width of $\sim 2.0$ (in the natural logarithm) depending on the particular
subsample. However, this was for all clumps, which inhabit clouds encompassing a 3 order of magnitude
range in mass. Since $s<1$, the mean clump mass increases for larger cloud masses, so the
clump mass function they observed is the convolution of the \MGMC\ distribution with
the conditional clump mass distribution $P(\Mclump|\MGMC)$, whose width is parametrized
by $\sigma_M$. Although we cannot directly derive $\sigma_M$ from those results, there is
one significant piece of evidence that suggests that $\sigma_M$ is not large: when
\citet{ellsworthbowers15} decompose their samples into ``Cloud'' vs. ``Clump/Core''
subsamples, the lognormal widths are indistinguishable within the errors, with
differences always smaller than $0.1$. It therefore appears that the width of the
clump mass distribution is driven mainly by the properties of the clouds they inhabit
rather than intrinsic clump-to-clump dispersion, and so I adopt $\sigma_M=0.1$ as the fiducial
model value.

\subsection{Mass-Radius Relation}
\citet{larson81}'s relations imply that molecular cloud masses and radii are related as
\begin{equation} M \propto R^{1.9} \end{equation}
or, surface density $\Sigma \equiv M / \pi R^2$ is approximately constant.
\citet{ellsworthbowers15} studied the mass-radius relation in BGPS clumps, and found that
$\Sigma \sim 100~\Msun~\mathrm{pc}^{-2}$ over five orders of magnitude in clump mass
with a slight increase at larger mass.
I therefore adopt a uniform value of $\Sigma_{\mathrm{clump}} = 100~\Msun~\mathrm{pc}^{-2}$
to determine clump radii from their mass. The exact form of the clump mass-radius relation
does not impact the final result because cross-enrichment dominates over self-enrichment
within a clump.

\subsection{Duration of Star Formation}
There is quite a bit of uncertainty in how long star formation occurs in a GMC.
Observations within the Milky Way and Local Group galaxies suggest GMC lifetimes of 20--30~Myr
\citep{blitz07,murray11}, corresponding to 2--3 free-fall times. Simulations of star formation
in GMCs show durations of star formation that range from $< 1$~Myr \citep{bate12}, to
``a few crossing times'', where $t_{\mathrm{cross}} \sim 10 (M / 10^6~\Msun)^{1/4}$~Myr
\citep{matzner02,krumholz06}, to $> 10$~free-fall times \citep{krumholztan07}. Looking at the
stellar populations of OB associations, Carina's star formation has proceeded for 3~Myr
and is still ongoing \citep{smithbrooks08}, providing a lower limit on the eventual duration of its star formation,
while the subgroups of Sco-Cen have internal age spreads of $1$--$7$~Myr, and average
ages between groups that spread by $6$~Myr \citep{pecaut12}. Gravitational timescales for
proto-GCs were presumably shorter than for Milky Way star formation regions due to their
compactness, adding further uncertainty. I adopt $\sigma_{\mathrm{tform}} = 10$~Myr for the
fiducial model.

\subsection{Mixing Efficiency}
Although \citetalias{bailin09} adopted the maximal mixing approximation (equivalent to $f_{\mathrm{mix}}=1$), outflows and simulations of starbursting dwarf galaxies find that their outflows contain most of the metals that were produced within the starburst \citep{martin02,roblesvaldez17}, suggesting that a large fraction of CCSN ejecta escape directly without mixing. However, no equivalent constraints exist on the metal enrichment of outflows from forming GC analogs, leaving $f_{\mathrm{mix}}$ relatively uncertain. Motivated by the dwarf galaxy studies, I adopt a small value of $f_{\mathrm{mix}}=0.2$ for the fiducial model.

\subsection{CCSN Yields}
Yields from CCSNe are quite uncertain; four examples of possible yields are given in Appendix~\ref{sec:yields}. Of these, the BH09, Nomoto13, and nuGrid models produce metals over a similar timescale but with a different total normalization, while the C\^ot\'e model produces significantly less metals and at significantly later times because it predicts that most high-mass stars collapse directly to a black hole instead of producing ejecta (Figure~\ref{fig:yield-functions}(b)).

Galactic chemical evolution sets constraints on nucleosynthetic yields \citep[e.g.][]{tinsley80}, with the details depending heavily on the assumptions about inflow, outflow, and the star formation history. A basic lower limit on yields can be obtained by assuming that gaseous inflow onto galaxies is relatively pristine; in such a case, the metal mass locked up in stars was all produced by the stellar population, and therefore
the total metal mass locked up in current stars ($= \left<Z_*\right> M_*$, where $\left<Z_{*}\right>$ is the mean stellar metallicity) must not be greater than amount of metals that could be produced by those stars ($= y M_*$, where $y$ is the total fraction of the mass of a stellar population that is returned to the ISM as metals, as plotted in Figure~\ref{fig:yield-functions}). In other words,
\begin{equation}\label{eq:yieldlimit-galCE}
  y \ge \left<Z_{*}\right>
\end{equation}

Although most work on the mass-metallicity relation of galaxies comes from gas phase metallicities \citep[e.g.][]{tremonti04}, which are higher than the mean stellar metallicities, stellar metallicities can be derived from detailed spectral modeling. For example, \citet{gallazzi05} found that there are substantial numbers of SDSS galaxies with metallicities as high as $2.5\times$ solar ($\log Z/Z_{\odot} = 0.4$), corresponding to a limit of $y \ge 0.04~M_{\odot}$. Conservatively, CCSNe should produce at least half of those heavy elements \citep[e.g.][who argue that Type Ia SNe contribute 10\%\ or less to the chemical enrichment of galaxies]{ferreras02}, meaning that the cumulative yield from CCSNe should be at least $0.02~M_{\odot}$. Comparing this to Figure~\ref{fig:yield-functions}, we see that the C\^ot\'e model does not satisfy this constraint.
I adopt the Nomoto13 yields, as the intermediate case among the three higher-yield models.

\subsection{Neglect of Other Energy Sources}
Although not truly a parameter, it is important to note that the fiducial model only includes the effects of CCSNe, and neglects stellar winds, which also produce nucleosynthetically-enriched material.
The total metallicity, and in particular the iron abundance, are indeed overwhelmingly dominated by CCSNe, and therefore the mass ejected by stellar winds can be neglected in this paper. However, stellar winds also inject energy into their surroundings, as does UV radiation from luminous massive stars. These sources could affect the energy budget of the gas, and therefore the ability of clumps and clouds to hold onto CCSN ejecta.

The effects of radiation on star forming clouds is predicted to dominate over that of stellar winds \citep{murray10}. The total amount of UV radiation produced by a stellar population is around $1$ -- $2 \times 10^{49}$~erg per $M_{\odot}$ of stars formed \citep{stinson13}, as much or more than produced by CCSNe ($\approx 1 \times 10^{49}$~erg). However, this energy is likely to be less efficient at coupling to gas ejection for two reasons: (1) many UV photons can directly escape the turbulent cloud without being absorbed \citep{howard17}; and (2) the heating is slow (unlike the rapid energy injection of supernovae), so cooling from dust and CO, which operate relatively quickly, have time to compensate and can efficiently radiate away large amounts of energy \citep{murray09}. Observationally, the amount of dust emission from super star clusters matches the bolometric luminosity of the massive stars reasonably well \citep{vanzi04}, suggesting that most of this energy is indeed lost.

In this paper, I therefore assume that the ejection of gas is dominated by the energy from CCSNe. It is, however, possible that this assumption underestimates the amount of gas that can be ejected.

\section{Results}\label{sec:results}

\subsection{Metallicity Distribution Functions}
The fundamental difference between the clumpy self-enrichment model and the original \citetalias{bailin09} model is that clusters can have internal metallicity distribution functions (MDFs); not all stars within the cluster have the same metallicity.

\begin{figure*}
\includegraphics[scale=0.6]{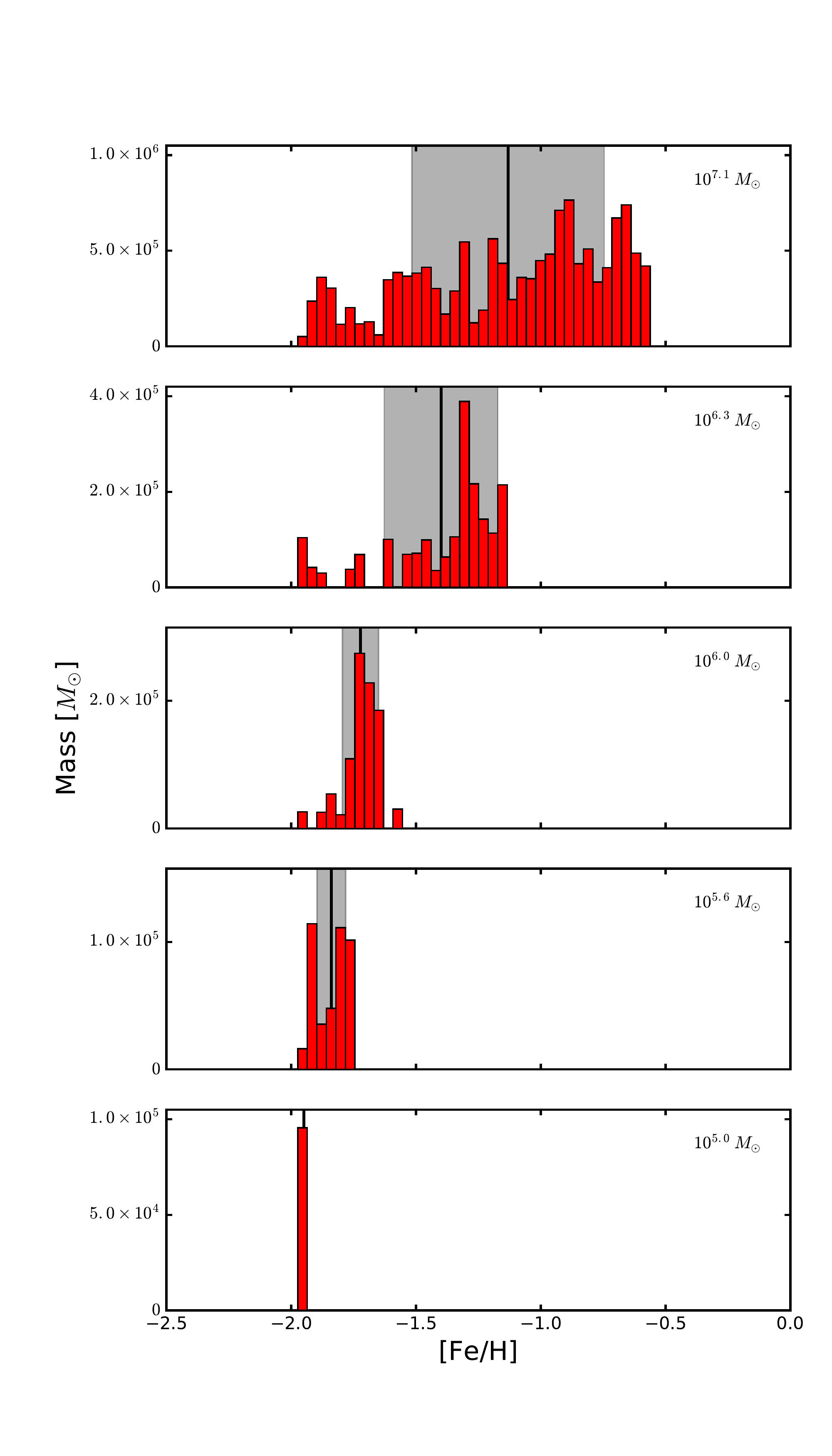}
\caption{\label{fig:histogram}%
From bottom to top, each histogram represents the metallicity distribution function (MDF) of the stars within
a GC model of increasing mass, labeled by GC stellar mass. The solid line shows the mean $\left<\mathrm{[Fe/H]}\right>$
of the cluster, while the gray band shows the $\pm1\sigma$ scatter derived from the histogram. The y-axes are in units of stellar mass per bin.
These particular clusters are highlighted with diamonds in Figures~\ref{fig:M_Z}--\ref{fig:Z_sigmaZ}.
}
\end{figure*}

Some example MDFs for model clusters with a range of final stellar masses are shown in Figure~\ref{fig:histogram}. The stellar mass of the model GC, indicated by the label on each histogram, is the main sequence mass of stars with lifetimes longer than 10~Gyr, plus a remnant mass (assumed to be $2~M_{\odot}$ on average for CCSN remnants and $0.56~M_{\odot}$ for white dwarfs) for shorter-lived stars.
For the lowest-mass clusters, all stars share the original pre-enrichment metallicity level of the cloud ($\mathrm{[Fe/H]}=-1.95$ in this case), but more massive clusters contain stars ranging from the pre-enrichment level and higher. As the mass of the GC increases, the number of clumps and the depth of the potential well increases, resulting in higher mean metallicities (indicated by the solid black line) and also wider MDFs (the gray shaded region indicates the $1\sigma$ ranges).

An interesting feature that appears naturally but not universally in the model is the appearance of distinct peaks in the MDFs; in Figure~\ref{fig:histogram}, this is particularly noticeable for the $10^{5.6}~M_{\odot}$, $10^{6.3}~M_{\odot}$, and $10^{7.1}~M_{\odot}$ clusters.
Observations of many of the classic iron-spread clusters also show distinct populations in total metallicity, such as in $\omega$~Cen \citep{johnson10}, M~22 \citep{marino09}, and Terzan~5 \citep{massari14}, though some show a more continuous distribution \citep{carretta10-m54gal}.
In the model, different random seeds result in different MDFs, but such peaks appear often, especially at these masses. These are each composed of a number of clumps, and appear entirely due to stochastic variation in the formation times of the clumps, rather than requiring a distinct formation mechanism for second-generation stars, as is required in many models of GC self-enrichment \citep[e.g.][]{dercole08,dantona16}.

One apparent difference between the shape of the model MDFs and those of observed iron-spread GCs is that the tallest peaks tend to occur at the lower metallicity end in observed GCs, but at the higher end for the model MDFs.

\subsection{Mass-Metallicity Relation (MMR)}

\begin{figure*}
\plottwo{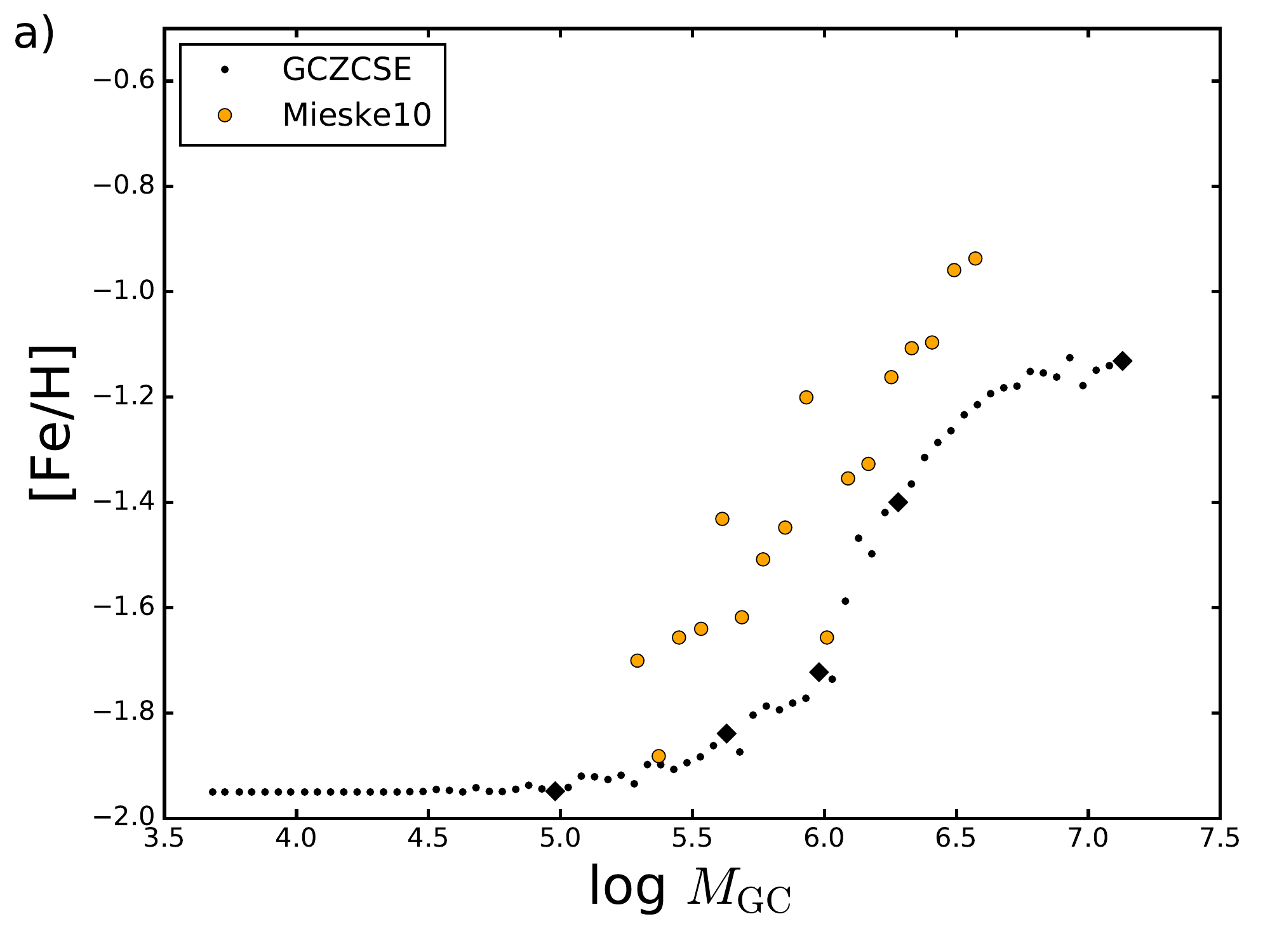}{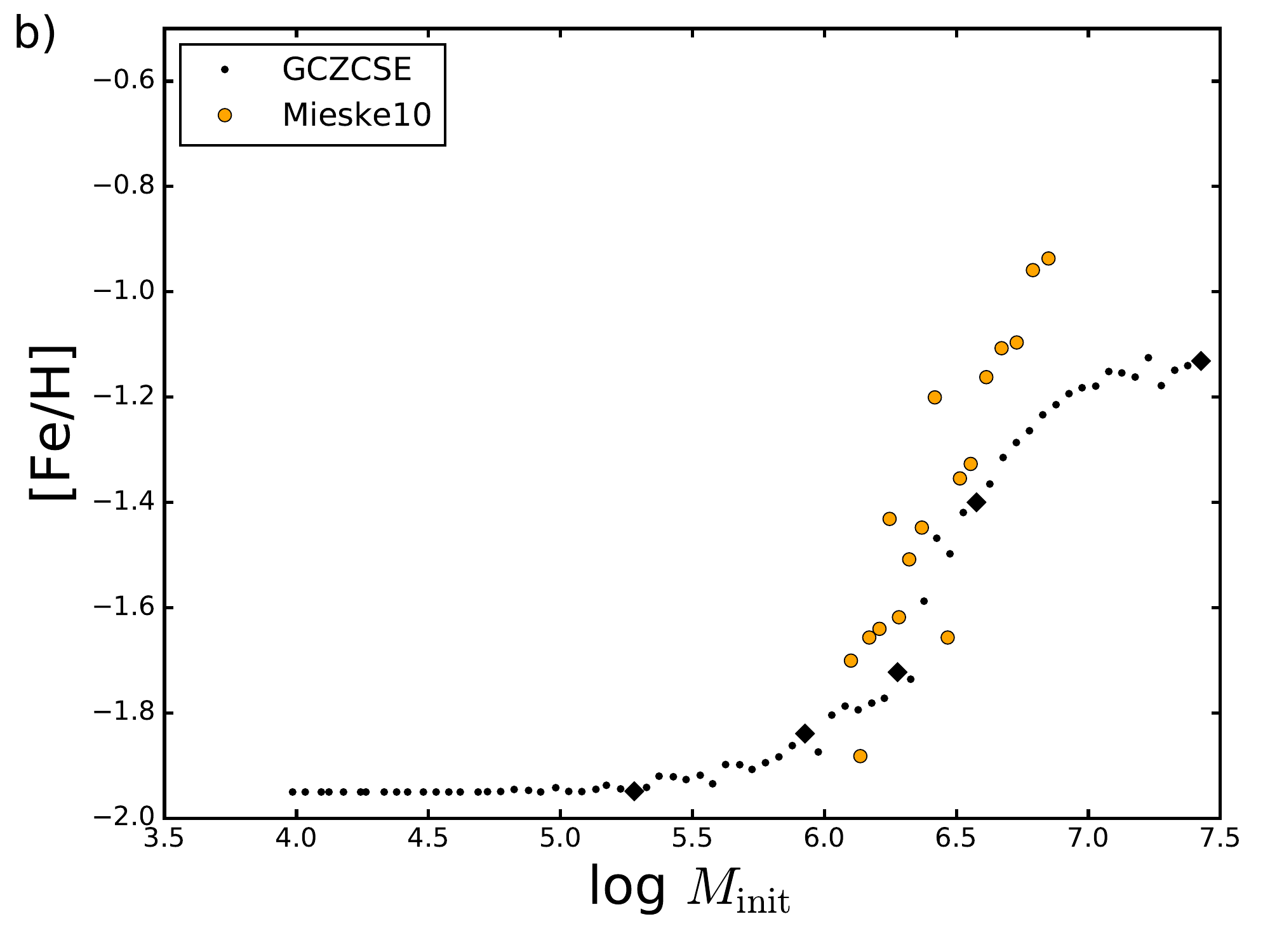}
\caption{\label{fig:M_Z}%
\textit{a)} Mean metallicity of GCs as a function of their current stellar mass for the fiducial clumpy self-enrichment model (black), and for the mean MMR among the metal-poor clusters around giant elliptical galaxies in the ACS Fornax Cluster Survey (\citealp{mieske10}; yellow). The black diamonds denote the particular model clusters whose MDFs are shown in Figure~\ref{fig:histogram}.
\textit{b)} As in panel (a), but using the initial stellar mass of the GC; for the observations, this is estimated using the mass loss prescription of \citet{goudfrooij14}.%
}
\end{figure*}

The net effect of the self-enrichment on clusters of a range of mass can be seen in Figure~\ref{fig:M_Z}. The left panel shows the MMR for clusters with a pre-enrichment level of $\mathrm{[Fe/H]}=-1.95$ compared to the blue tilt measurements of \citet{mieske10} based on GC systems around ellipticals in the Fornax cluster. Diamonds denote the particular model clusters whose MDFs are shown in Figure~\ref{fig:histogram}.
The model predicts the observed blue tilt qualitatively, preserving the success of the single-zone \citetalias{bailin09} model.

The predicted blue tilt begins at higher mass than observed by a factor of $\sim 2$. However, the model GCs only lose mass due to stellar evolution, while observed GCs also lose mass due to dynamical evolution. Therefore, it is better to compare the clusters according to their initial stellar mass; for the observations, I use the mean relation between initial and final cluster masses from figure 5b of \citet{goudfrooij14}.
This is shown in Figure~\ref{fig:M_Z}(b) and the quantitative agreement is good.

The effects of self-enrichment appear slowly at low masses, then increase sharply around $M_{\mathrm{GC}} \sim 10^{6.0}~M_{\odot}$ or $M_{\mathrm{init}} \sim 10^{6.3}~M_{\odot}$. This is a consequence of mixing, in particular the transition in Equation~(\ref{eq:mixingcloud}) between the two terms in the minimum. In other words, below this mass, the unmixed component of the ejecta is lost completely, so only the mixed component contributes to self-enrichment, while above this mass, the wind is not strong enough to completely remove the unmixed metals, so some of them are available for self-enrichment. The sharpness of the transition is an artifact of the way the model is constructed, but the underlying behavior is physical.

\subsection{Internal Metallicity Spreads}\label{sec:results-sigma}

\begin{figure*}
\plottwo{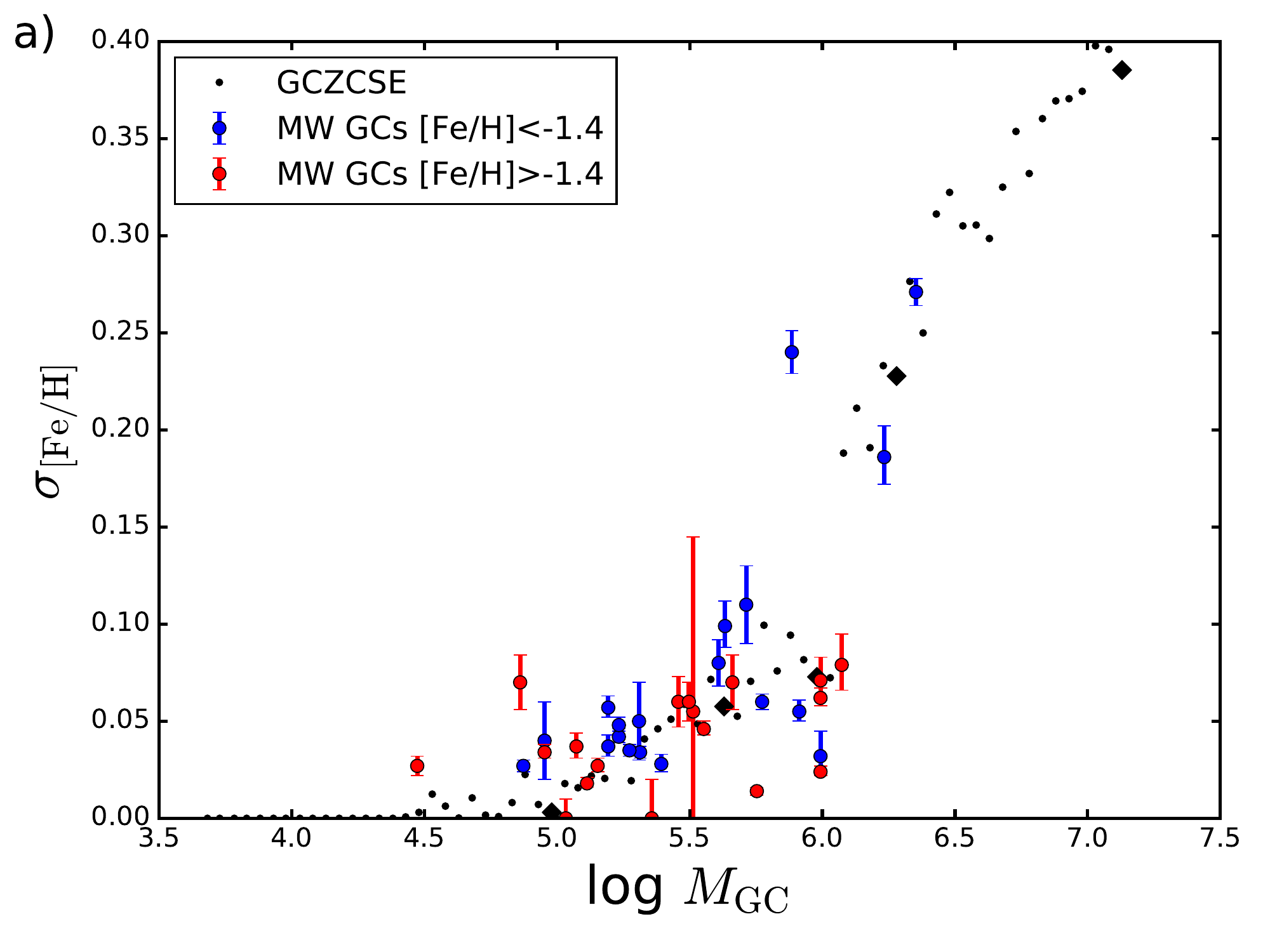}{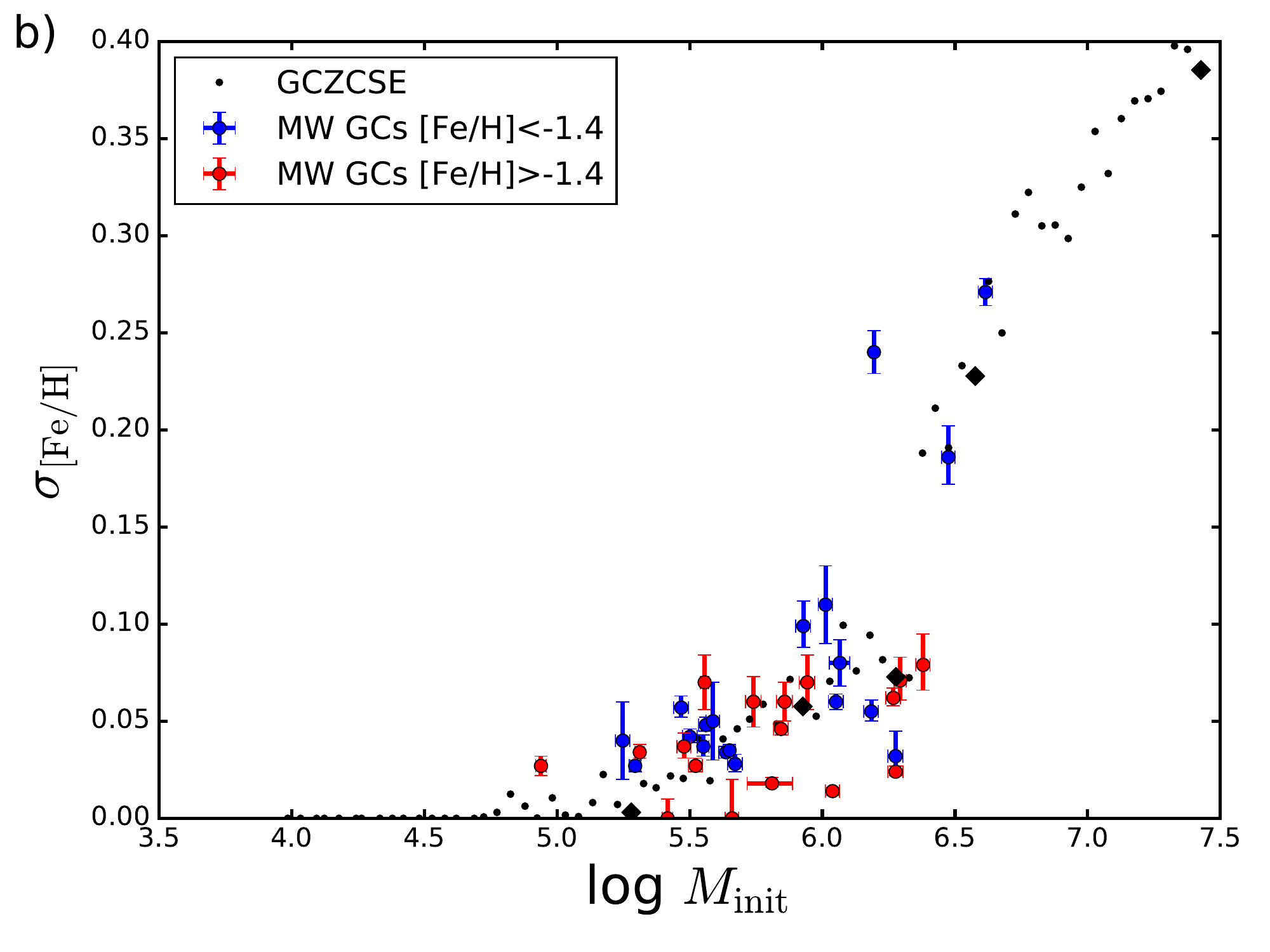}
\caption{\label{fig:M_sigmaZ}%
\textit{a)} Internal metallicity spread $\sigma_{\mathrm{[Fe/H]}}$ of GCs as a function of their current stellar mass for the fiducial clumpy self-enrichment model (black) and for observed Milky Way metal-poor (blue) and metal-rich (red) clusters. The black diamonds denote the particular model clusters whose MDFs are shown in Figure~\ref{fig:histogram}. Observations are drawn from the literature (see Section~\ref{sec:results-sigma} for details).
\textit{b)} As in panel (a), but plotting each observed GC at its initial stellar mass. For the observed GCs, the estimated initial mass accounting for stellar evolution and tidal stripping, where available, is taken from \citet{balbinot18}.%
}
\end{figure*}

\begin{figure}
\plotone{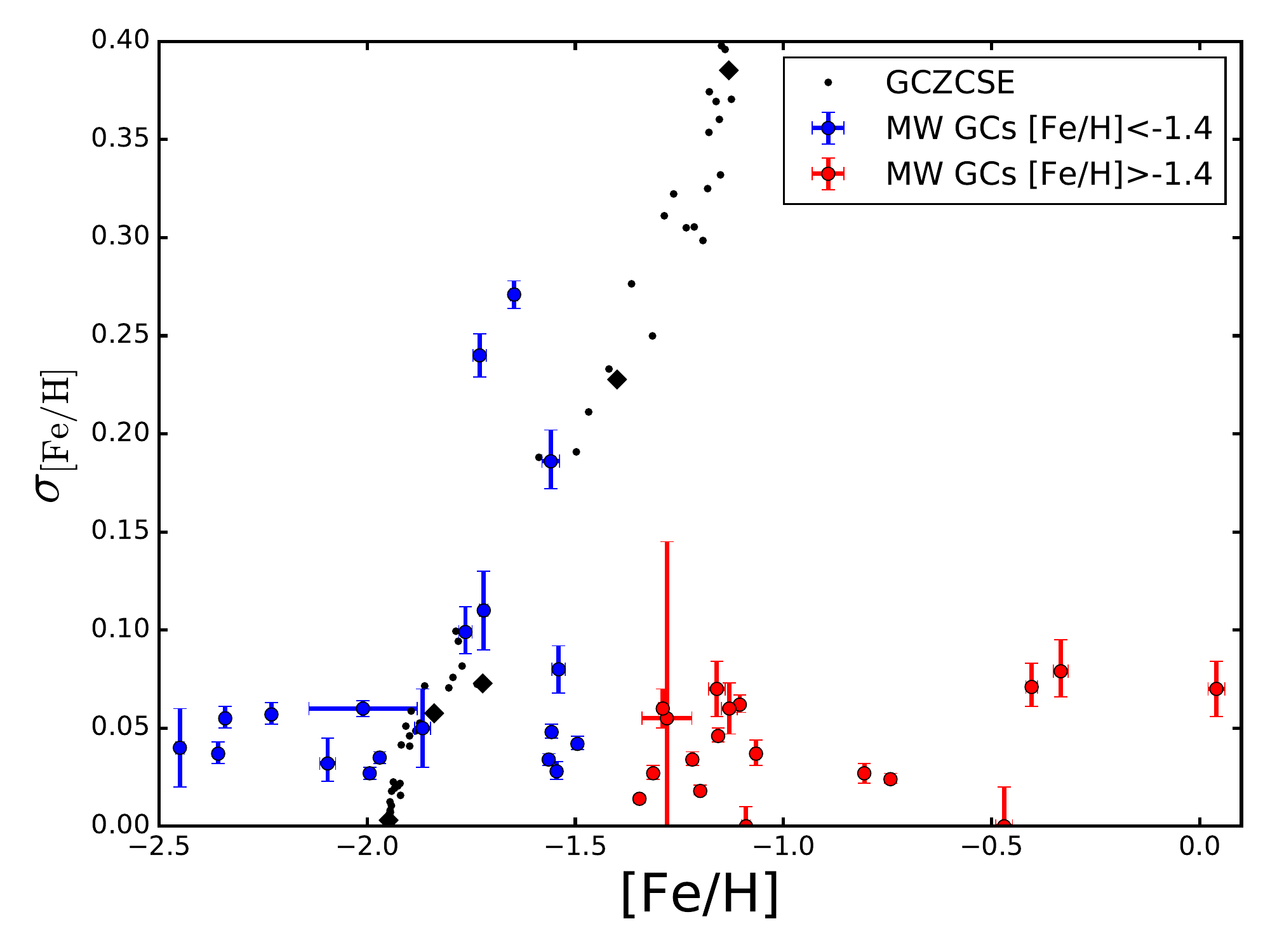}
\caption{\label{fig:Z_sigmaZ}%
Internal metallicity spread as a function of mean metallicity, for model clusters (black) and observed Milky Way clusters (blue and red, as in Figure~\ref{fig:M_sigmaZ}). The black diamonds denote the particular model clusters whose MDFs are shown in Figure~\ref{fig:histogram}. Clusters that formed from clouds of the same composition (in this case, [Fe/H] = $-1.95$) but that experienced different levels of self-enrichment occupy a diagonal sequence; for example, the four Milky Way clusters at [Fe/H] = $-1.87$ (NGC~5634), $-1.76$ (M~22), $-1.72$ (NGC~5286), and $-1.56$ (M~54) are all consistent with forming from the same material.%
}
\end{figure}

The most important prediction of this model is that
the internal metallicity spread depends on cluster mass. This is shown in Figure~\ref{fig:M_sigmaZ}. At masses $M_{\mathrm{GC}} \lesssim 10^{5.0}~M_{\odot}$, there is no internal spread, but the predicted spread increases steadily up to $M_{\mathrm{GC}} \sim 10^{6.0}~M_{\odot}$, at which point it increases sharply due to the mixing prescription, as described above,
 until $M_{\mathrm{GC}} \sim 10^{6.5}~M_{\odot}$, where it saturates.
The saturation occurs as the number of clumps becomes very large, and the chemical enrichment resembles an analytic leaky box model with a broad peaked distribution \citep[e.g.][]{prantzos08}.
The GCZCSE model therefore predicts that GCs should have internal spreads in elements produced by CCSNe, especially for high masses $M_{\mathrm{GC}} \gtrsim 10^{5.5}~M_{\odot}$.

The common lore is that GCs do \textbf{not} have an internal spread in iron or other elements that come predominantly from supernovae except for a few anomalous clusters that might actually be stripped nuclei of dwarf galaxies \citep{bekkifreeman03,carretta10-m54gal,massari14}. However, the observational data themselves show a more nuanced story. Observations of internal iron spreads are plotted in Figure~\ref{fig:M_sigmaZ}(a) in blue and red for metal-poor and metal-rich clusters respectively, from the compilation of \citet{willmanstrader12} supplemented by more recent observations \citep{kacharov13,dacosta14,boberg15,marino15,dalessandro16,mucciarelli16,carretta17,johnson17-m19,johnson17-ngc5986,johnson17-ngc6229,liu17,villanova17,muraguzman18}. Most clusters show internal spreads of $\sigma_{\mathrm{[Fe/H]}} < 0.1$, which is indeed small, but in most cases statistically inconsistent with zero; more interestingly, the observations show an increase in $\sigma_{\mathrm{[Fe/H]}}$ with mass, as predicted \citep[see also][]{willmanstrader12,dacosta16}. The quantitative agreement is fair, but, as for the MMR, the model predicts that the increase in iron spread begins at higher mass than observed.

For Milky Way GCs, the amount of mass loss from stellar evolution and tidal stripping has been estimated by \citet{balbinot18} on a cluster-by-cluster basis, allowing us to directly compare the model and observed clusters using their initial masses, shown Figure~\ref{fig:M_sigmaZ}(b). The quantitative match between the predictions and observations is excellent. This is particularly impressive since the parameters of the model are tuned based on the properties of local star-forming regions (Section~\ref{sec:fiducial-parameters}) not on GCs. This model successfully predicts both the spreads in typical GCs, and also the large spreads seen in anomalous clusters such as $\omega$~Cen, M~54, and NGC~6273 (the points at $\sigma_{\mathrm{[Fe/H]}}=0.27$, $0.19$, and $0.24$ respectively). In other words, the internal spread in iron seen in these clusters is \textit{completely consistent} with being formed from the same physics as other GCs and does not require them to be stripped dwarf galaxy nuclei\footnote{I am not arguing that they \textit{cannot} be stripped dwarfs, merely that such a scenario is \textit{not required} to explain the widths of their MDFs.}.

Finally, we can directly see the effects of clumpy self-enrichment by plotting mean metallicity [Fe/H] vs. internal metallicity spread $\sigma_{\mathrm{[Fe/H]}}$ (Figure~\ref{fig:Z_sigmaZ}). A series of models are shown that all come from a gas cloud pre-enriched to a level of $\mathrm{[Fe/H]}=-1.95$ but with different masses, and therefore different amounts of self-enrichment. In the GCZCSE model, self-enrichment both increases the mean metallicity, moving them to the right, and also widens the MDF, moving them upward. Therefore, clusters that formed from clouds of the same composition but which experienced different levels of self-enrichment occupy a diagonal sequence with an $x$-intercept at the metallicity of the original gas cloud; for example, the four Milky Way clusters at [Fe/H] = $-1.87$ (NGC~5634), $-1.76$ (M~22), $-1.72$ (NGC~5286), and $-1.56$ (M~54), which lie along the plotted model points, are all consistent with coming from an $\mathrm{[Fe/H]}=-1.95$ gas cloud.

\subsection{Sensitivity to Model Parameters}\label{sec:parameter-dependence}
\begin{figure*}
\plotone{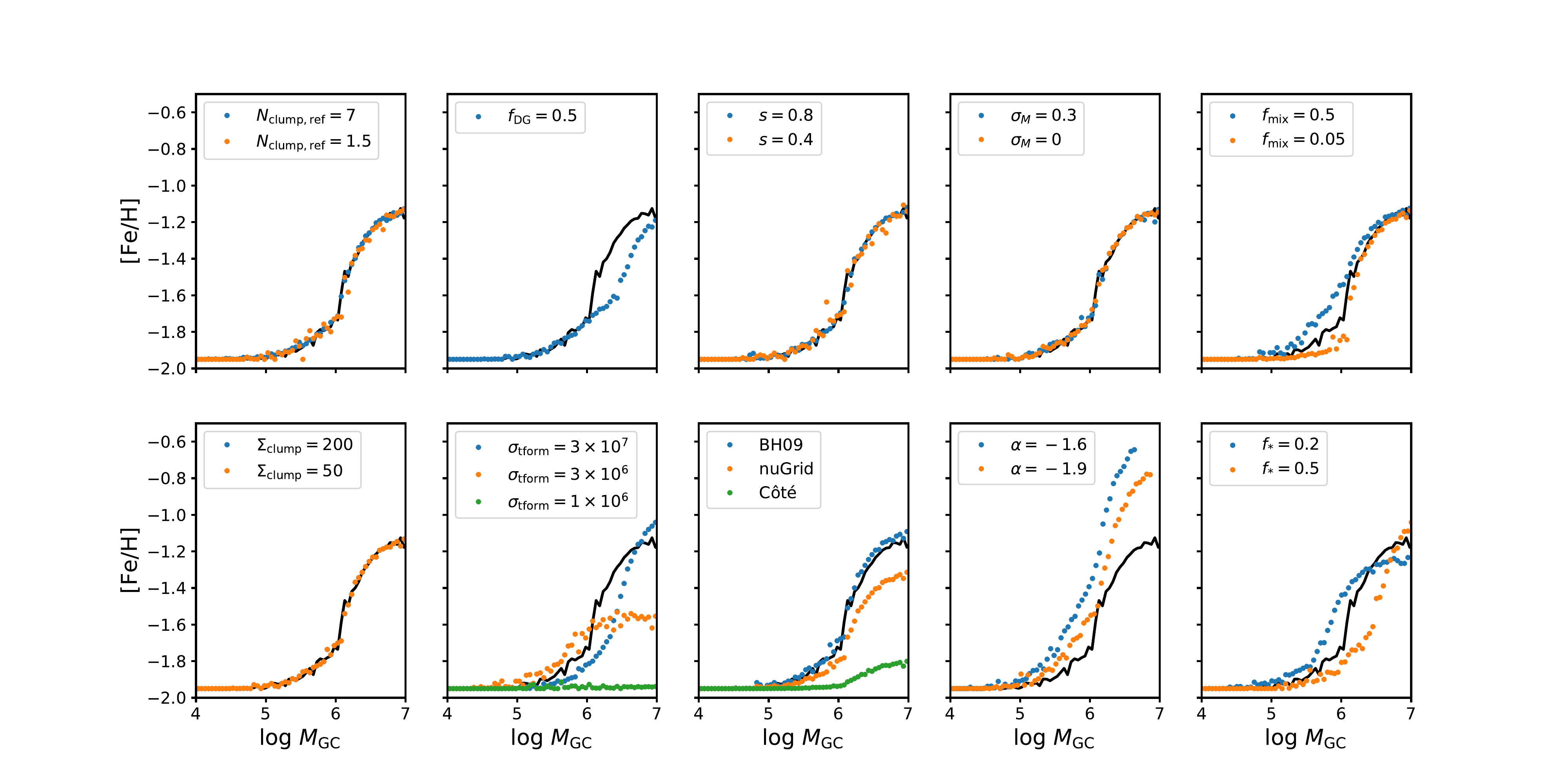}
\caption{\label{fig:param-mz}%
Mass-metallicity relation, as in Figure~\ref{fig:M_Z}, for different choices of model parameters. In each case, the solid blank line is for the fiducial parameters (Table~\ref{table:fiducial parms}). Note that  $\Sigma_{\mathrm{clump}}$ has no effect, so the different model points all lie on top of each other.%
}
\end{figure*}

\begin{figure*}
\plotone{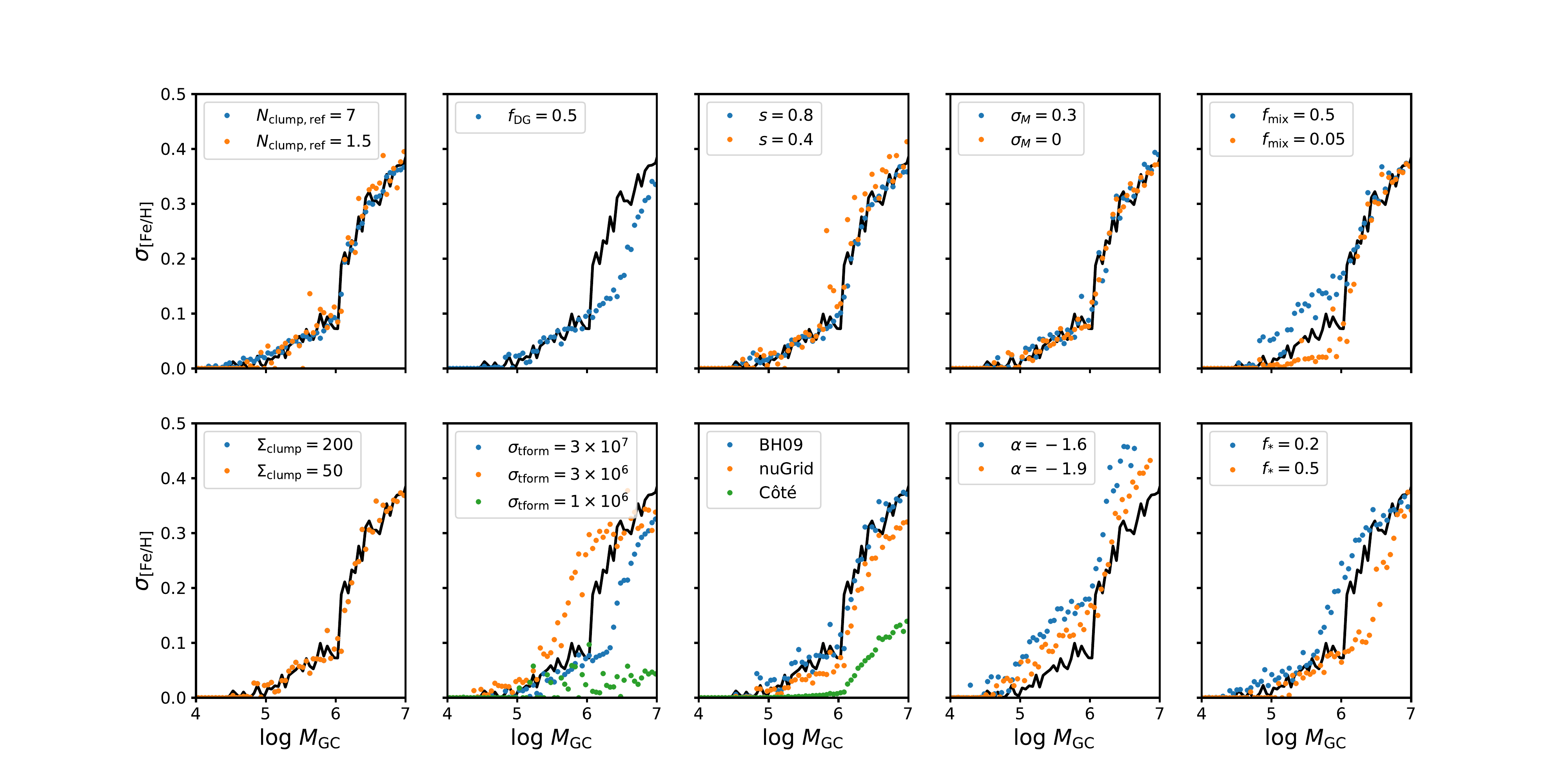}
\caption{\label{fig:param-msigma}%
Relation between cluster mass and internal metallicity spread, as in Figure~\ref{fig:M_sigmaZ}, for different choices of model parameters. In each case, the solid blank line is for the fiducial parameters (Table~\ref{table:fiducial parms}). Note that  $\Sigma_{\mathrm{clump}}$ has no effect, so the different model points all lie on top of each other.%
}
\end{figure*}

\begin{figure*}
\plotone{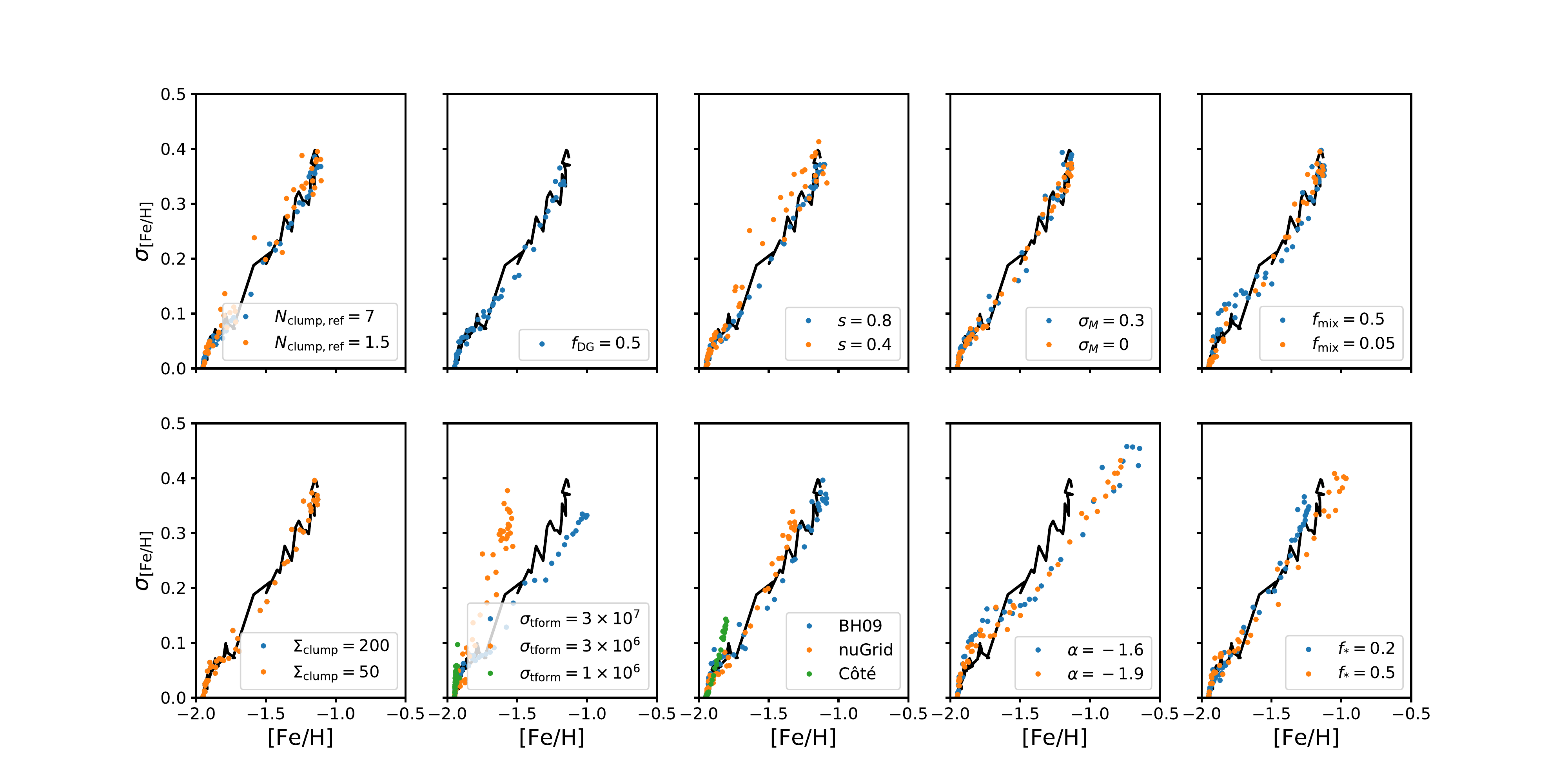}
\caption{\label{fig:param-zsigma}%
Relation between mean cluster metallicity, [Fe/H], and internal metallicity spread, $\sigma_{\mathrm{[Fe/H]}}$, as in Figure~\ref{fig:Z_sigmaZ}, for different choices of model parameters. In each case, the solid blank line is for the fiducial parameters (Table~\ref{table:fiducial parms}). Note that  $\Sigma_{\mathrm{clump}}$ has no effect, so the different model points all lie on top of each other. This relationship is a \textbf{remarkably robust} prediction of the model, independent of the model parameters. %
}
\end{figure*}

Figures~\ref{fig:param-mz}, \ref{fig:param-msigma}, and \ref{fig:param-zsigma} demonstrate how the model predictions depend on the parameters of the model. It is apparent that the general trends of the relations --- metal enhancement and metallicity dispersion that slowly increases with mass, then rise more sharply before saturating at high mass --- are generic predictions of the model.
Also, the form of the mass-radius relation of clumps and all of the clumping parameters have very little effect on the model, except for $f_{\mathrm{DG}}$, which can slightly affect the amount of self-enrichment for the most massive GCs.
The mixing efficiency has a modest effect on how much enrichment occurs at lower cluster masses, where the maximal mixing model retains more metals than the minimal mixing case, but has no influence at higher mass where the self-enrichment saturates.

\begin{figure}
\plotone{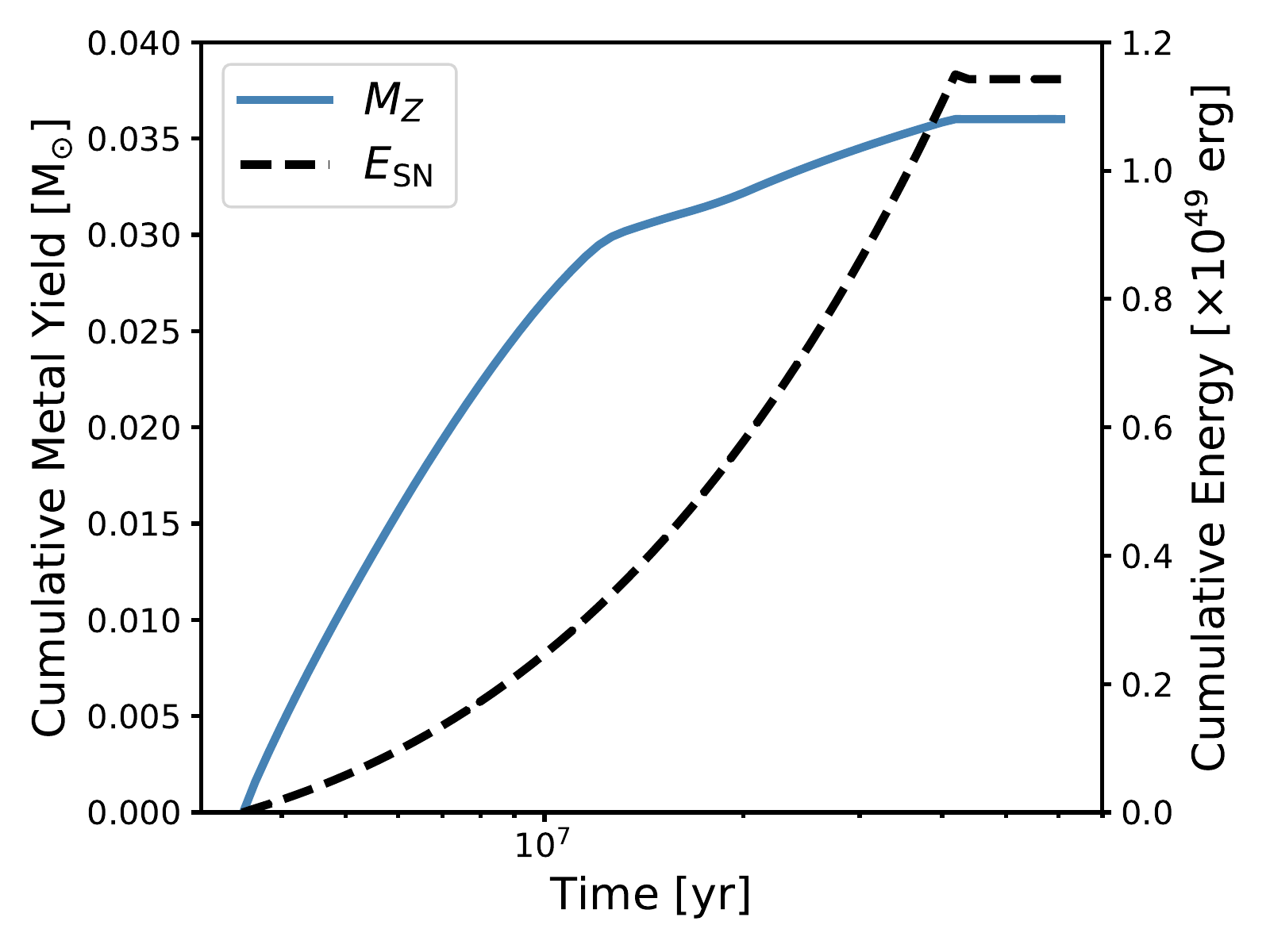}
\caption{\label{fig:MZESN}%
Cumulative production of metals (blue line, left-hand axis) and energy (black dashed line, right-hand axis) from CCSNe per solar mass of stars formed, as a function of time for the truncated Salpeter IMF and the Nomoto13 yields. Most of the metals are produced during the first few Myr, while most of the energy is released after $\sim 15$~Myr.%
}
\end{figure}

 The timescale of star formation, $\sigma_{\mathrm{tform}}$ has a significant influence on self-enrichment. For very short timescales ($\sigma_{\mathrm{tform}}=1\times 10^6$~yr), most clumps complete their star formation before any CCSNe have gone off, and so there is virtually no enrichment. However, once the star formation timescale approaches the delay before the first CCSNe occur, a significant amount of clumpy self-enrichment occurs. Interestingly, lower-mass clusters have the most self-enrichment if the duration of star formation is somewhat short ($<10^7$~yr), while higher-mass clusters have more self-enrichment if $\sigma_{\mathrm{tform}}$ is long. The reason for this is that higher-mass progenitor stars have larger relative metal yields, while all CCSNe release the same amount of energy; therefore, the release of metals happens faster than the release of energy (Figure~\ref{fig:MZESN}). Low-mass clusters, with shallow potential wells, are more sensitive to the amount of energy --- if the time delay between clumps is too long, the extra energy pushes most of the material out of the cloud, so it is not available for enrichment. For higher-mass clusters with deep potential wells, most of the material remains bound regardless of the amount of energy, so longer time delays between clumps allows time for more metals to be released, and therefore more self-enrichment.

The C\^ot\'e yields, which do not produce any CCSNe within the first 10~Myr, produce virtually no self-enrichment; however, these yields are already inconsistent with galactic chemical enrichment constraints. The BH09 (nuGrid) yields, which produce slightly more (less) metals but over a similar timescale as the Nomoto13 yields, have total metallicities and metallicity spreads that are correspondingly slightly higher (lower) than for the Nomoto13 yields.

The IMF also affects both the amount of self enrichment and the metallicity spread.  Top-heavier IMFs, which both have more CCSNe and more CCSNe from high mass stars (which have higher proportional yields), self-enrich to higher levels. Changing the star formation efficiency changes the mapping between GC stellar mass and the depth of the cloud potential well, and so larger values of $f_*$ correspond to less efficient self-enrichment due to shallower potential wells at a given value of \MGC.

 The most interesting result of the parameter study is that
 the relationship between metallicity, [Fe/H], and internal metallicity spread, $\sigma_{\mathrm{[Fe/H]}}$ in Figure~\ref{fig:param-zsigma} is a \textbf{remarkably robust} prediction of the model, mostly independent of the model parameters! In other words, the internal spread $\sigma_{\mathrm{[Fe/H]}}$ is an excellent measurement of how much self-enrichment has occurred. The only parameter that has any measurable influence is $\sigma_{\mathrm{tform}}$, in the sense that short timescales have less total self-enrichment for a given value of $\sigma_{\mathrm{[Fe/H]}}$.

\section{Discussion and Conclusions}\label{sec:conclusions}
This paper presents a model for clumpy self-enrichment by CCSNe during the formation of GCs. The model is motivated by observations and simulations of star-forming regions in the Milky Way, in which GMCs fragment into clumps which each form stars at slightly different times and interact via gas flowing between clumps, before finally merging to form the cluster that we see today, 10~Gyr later.

The model predicts that more massive clusters, which have deeper potential wells and are therefore able to hold onto more of their CCSN ejecta, experience more self-enrichment, resulting in a mass-metallicity relationship that matches the observed ``blue tilt'' seen in the GC populations around massive elliptical galaxies, once mass loss is taken into account. It also predicts that more massive clusters have subpopulations with a broader range of metallicities, also in good agreement with observations of Milky Way GCs.

The conventional wisdom is that most GCs have ``no'' internal spread in iron abundance, but observations actually show a non-zero spread that increases with mass, in excellent agreement with the model. Observations often show distinct subpopulations, which also appear in the model mainly due to stochasticity. Interestingly, even those GCs with very large internal metallicity spreads (e.g. $\omega$~Cen, M~54, NGC~6273), which are sometimes considered to be the stripped nuclei of dwarf galaxies in part \textit{due} to the iron abundances, are matched very well by the model --- GC formation \textbf{should} produce clusters of that mass with internal iron abundance spreads that large, according to the clumpy self-enrichment model.

Because self-enrichment both increases the total metallicity of the cluster and creates internal metallicity spreads, the internal spread in iron abundances is an excellent measurement of the degree to which self-enrichment has modified the original abundances. This relationship is remarkably robust to changes in the model parameters, which opens up the possibility of using metallicity dispersions to reconstruct the metallicity of the original protocluster cloud. This idea is developed further in Bailin (in prep.).

A key assumption in this model is that GC formation is not fundamentally different from star formation that occurs today in the Milky Way, and in particular that scaling relations derived from the fragmentation of Milky Way GMCs can be applied to the much higher-density gas clouds at higher redshift in which GCs must have formed. In the model, parameters related to fragmentation affect the onset mass at which self-enrichment becomes efficient, but do not fundamentally change the results. The good match between the fiducial model and the observations can therefore be taken as evidence (or at least a consistency check) that these scaling relations may indeed extend to much more intense star formation events.

The model predictions are relatively insensitive to the model parameters. The duration of star formation is the one parameter that changes the form of the predicted relations, while the form of the IMF and the mixing efficiency of supernova ejecta can also increase or decrease the amount and variability of enrichment at certain cluster masses.

It is important to stress the the model presently only accounts for the influence of CCSNe. The neglect of heating from UV radiation of massive luminous stars could result in an overestimate on the amount of self-enrichment, though this energy source is likely less efficient than supernovae at driving gas ejection.
Also, the model only predicts the \textit{total} metallicity of stars, as traced by [Fe/H]. On the other hand, much of the observational evidence for self-enrichment and for multiple populations within clusters comes from light elements such as O and Na, which probably require a different enrichment source such as AGB winds, fast rotating massive stars, binary stars, or very massive stars \citep{dercole08,bastian18}. The code that instantiates GCZCSE is designed so that it can be extended to follow elements individually, and to include multiple sources of enrichment and of energetic feedback; this is the topic of future work. However, it is fair to ask whether the multiple populations seen in light elements and in iron are of fundamentally different origin, given that the enrichment sources are different. \citet{renzini13} argues that there is no fundamental difference, and in GCZCSE, this is true ---
although the enrichment in the present model comes entirely from CCSNe, the multiple populations arise fundamentally because of the clumpiness of star formation. This is in contrast to most theoretical models of GC formation, where multiple populations come from different episodes of star formation that may be well separated in time (e.g. $10^8$~yr) but not in space \citep{dantona16}, even when they consider a fractally fragmented gas cloud \citep{bekki17}. Whether this will work out quantitatively for other elemental abundances remains to be seen.

\acknowledgments

Thanks to Bill Harris, Dave Arnett, Yancy Shirley, Dean Townsley, and 
members of the Facebook group \textit{Python users in Astronomy}
for useful conversations. Thanks to Chris West and Alexander Heger for access to pre-publication versions of their supernova yields. This research made use of Astropy, a community-developed core Python package for Astronomy \citep{astropy}. 

Support for program HST-AR-13908.001-A was provided by NASA through a grant from the
Space Telescope Science Institute, which is operated by the Association for Research
in Astronomy, Inc., under NASA contract NAS 5-26555.

\bibliography{gczcseref.bib}

\begin{thebibliography}{}
\expandafter\ifx\csname natexlab\endcsname\relax\def\natexlab#1{#1}\fi

\bibitem[{{Alves} {et~al.}(2007){Alves}, {Lombardi}, \& {Lada}}]{alves07}
{Alves}, J., {Lombardi}, M., \& {Lada}, C.~J. 2007, \aap, 462, L17

\bibitem[{{Astropy Collaboration} {et~al.}(2013){Astropy Collaboration},
  {Robitaille}, {Tollerud}, {Greenfield}, {Droettboom}, {Bray}, {Aldcroft},
  {Davis}, {Ginsburg}, {Price-Whelan}, {Kerzendorf}, {Conley}, {Crighton},
  {Barbary}, {Muna}, {Ferguson}, {Grollier}, {Parikh}, {Nair}, {Unther},
  {Deil}, {Woillez}, {Conseil}, {Kramer}, {Turner}, {Singer}, {Fox}, {Weaver},
  {Zabalza}, {Edwards}, {Azalee Bostroem}, {Burke}, {Casey}, {Crawford},
  {Dencheva}, {Ely}, {Jenness}, {Labrie}, {Lim}, {Pierfederici}, {Pontzen},
  {Ptak}, {Refsdal}, {Servillat}, \& {Streicher}}]{astropy}
{Astropy Collaboration}, {Robitaille}, T.~P., {Tollerud}, E.~J., {et~al.} 2013,
  \aap, 558, A33

\bibitem[{{Bailin}(2018)}]{bailin18-gczcse-software}
{Bailin}, J. 2018, {GCZCSE}, v.1.0.0,  Zenodo, doi:10.5281/zenodo.1253347

\bibitem[{{Bailin} \& {Harris}(2009)}]{bailin09}
{Bailin}, J., \& {Harris}, W.~E. 2009, \apj, 695, 1082

\bibitem[{{Balbinot} \& {Gieles}(2018)}]{balbinot18}
{Balbinot}, E., \& {Gieles}, M. 2018, \mnras, 474, 2479

\bibitem[{{Bastian} \& {Lardo}(2018)}]{bastian18}
{Bastian}, N., \& {Lardo}, C. 2018, \araa, 56, in press, arXiv:1712.01286

\bibitem[{{Bate}(2009)}]{bate09}
{Bate}, M.~R. 2009, \mnras, 392, 590

\bibitem[{{Bate}(2012)}]{bate12}
---. 2012, \mnras, 419, 3115

\bibitem[{{Battisti} \& {Heyer}(2014)}]{battisti14}
{Battisti}, A.~J., \& {Heyer}, M.~H. 2014, \apj, 780, 173

\bibitem[{{Bekki}(2017)}]{bekki17}
{Bekki}, K. 2017, \mnras, 469, 2933

\bibitem[{{Bekki} \& {Freeman}(2003)}]{bekkifreeman03}
{Bekki}, K., \& {Freeman}, K.~C. 2003, \mnras, 346, L11

\bibitem[{{Blitz} {et~al.}(2007){Blitz}, {Fukui}, {Kawamura}, {Leroy},
  {Mizuno}, \& {Rosolowsky}}]{blitz07}
{Blitz}, L., {Fukui}, Y., {Kawamura}, A., {et~al.} 2007, Protostars and Planets
  V, 81

\bibitem[{{Boberg} {et~al.}(2015){Boberg}, {Friel}, \& {Vesperini}}]{boberg15}
{Boberg}, O.~M., {Friel}, E.~D., \& {Vesperini}, E. 2015, \apj, 804, 109

\bibitem[{{Carretta} {et~al.}(2017){Carretta}, {Bragaglia}, {Lucatello},
  {D'Orazi}, {Gratton}, {Donati}, {Sollima}, \& {Sneden}}]{carretta17}
{Carretta}, E., {Bragaglia}, A., {Lucatello}, S., {et~al.} 2017, \aap, 600,
  A118

\bibitem[{{Carretta} {et~al.}(2009){Carretta}, {Bragaglia}, {Gratton},
  {Lucatello}, {Catanzaro}, {Leone}, {Bellazzini}, {Claudi}, {D'Orazi},
  {Momany}, {Ortolani}, {Pancino}, {Piotto}, {Recio-Blanco}, \&
  {Sabbi}}]{carretta09}
{Carretta}, E., {Bragaglia}, A., {Gratton}, R.~G., {et~al.} 2009, \aap, 505,
  117

\bibitem[{{Carretta} {et~al.}(2010){Carretta}, {Bragaglia}, {Gratton},
  {Lucatello}, {Bellazzini}, {Catanzaro}, {Leone}, {Momany}, {Piotto}, \&
  {D'Orazi}}]{carretta10-m54gal}
---. 2010, \apjl, 714, L7

\bibitem[{{Chabrier}(2003)}]{chabrier03}
{Chabrier}, G. 2003, \pasp, 115, 763

\bibitem[{{Choi} {et~al.}(2016){Choi}, {Dotter}, {Conroy}, {Cantiello},
  {Paxton}, \& {Johnson}}]{choi16}
{Choi}, J., {Dotter}, A., {Conroy}, C., {et~al.} 2016, \apj, 823, 102

\bibitem[{{C{\^o}t{\'e}} {et~al.}(2016){C{\^o}t{\'e}}, {West}, {Heger},
  {Ritter}, {O'Shea}, {Herwig}, {Travaglio}, \& {Bisterzo}}]{cote16}
{C{\^o}t{\'e}}, B., {West}, C., {Heger}, A., {et~al.} 2016, \mnras, 463, 3755

\bibitem[{{da Costa}(2016)}]{dacosta16}
{da Costa}, G.~S. 2016, in IAU Symposium, Vol. 317, The General Assembly of
  Galaxy Halos: Structure, Origin and Evolution, ed. A.~{Bragaglia},
  M.~{Arnaboldi}, M.~{Rejkuba}, \& D.~{Romano}, 110--115

\bibitem[{{Da Costa} {et~al.}(2014){Da Costa}, {Held}, \&
  {Saviane}}]{dacosta14}
{Da Costa}, G.~S., {Held}, E.~V., \& {Saviane}, I. 2014, \mnras, 438, 3507

\bibitem[{{Dalessandro} {et~al.}(2016){Dalessandro}, {Lapenna}, {Mucciarelli},
  {Origlia}, {Ferraro}, \& {Lanzoni}}]{dalessandro16}
{Dalessandro}, E., {Lapenna}, E., {Mucciarelli}, A., {et~al.} 2016, \apj, 829,
  77

\bibitem[{{D'Antona} {et~al.}(2016){D'Antona}, {Vesperini}, {D'Ercole},
  {Ventura}, {Milone}, {Marino}, \& {Tailo}}]{dantona16}
{D'Antona}, F., {Vesperini}, E., {D'Ercole}, A., {et~al.} 2016, \mnras, 458,
  2122

\bibitem[{{D'Ercole} {et~al.}(2008){D'Ercole}, {Vesperini}, {D'Antona},
  {McMillan}, \& {Recchi}}]{dercole08}
{D'Ercole}, A., {Vesperini}, E., {D'Antona}, F., {McMillan}, S.~L.~W., \&
  {Recchi}, S. 2008, \mnras, 391, 825

\bibitem[{{Ellsworth-Bowers} {et~al.}(2015){Ellsworth-Bowers}, {Glenn},
  {Riley}, {Rosolowsky}, {Ginsburg}, {Evans}, {Bally}, {Battersby}, {Shirley},
  \& {Merello}}]{ellsworthbowers15}
{Ellsworth-Bowers}, T.~P., {Glenn}, J., {Riley}, A., {et~al.} 2015, \apj, 805,
  157

\bibitem[{{Ferreras} \& {Silk}(2002)}]{ferreras02}
{Ferreras}, I., \& {Silk}, J. 2002, \mnras, 336, 1181

\bibitem[{{Gallazzi} {et~al.}(2005){Gallazzi}, {Charlot}, {Brinchmann},
  {White}, \& {Tremonti}}]{gallazzi05}
{Gallazzi}, A., {Charlot}, S., {Brinchmann}, J., {White}, S.~D.~M., \&
  {Tremonti}, C.~A. 2005, \mnras, 362, 41

\bibitem[{{Goudfrooij} \& {Kruijssen}(2014)}]{goudfrooij14}
{Goudfrooij}, P., \& {Kruijssen}, J.~M.~D. 2014, \apj, 780, 43

\bibitem[{{Harris} {et~al.}(2006){Harris}, {Whitmore}, {Karakla}, {Oko{\'n}},
  {Baum}, {Hanes}, \& {Kavelaars}}]{harris06}
{Harris}, W.~E., {Whitmore}, B.~C., {Karakla}, D., {et~al.} 2006, \apj, 636, 90

\bibitem[{{Howard} {et~al.}(2017){Howard}, {Pudritz}, \& {Klessen}}]{howard17}
{Howard}, C., {Pudritz}, R., \& {Klessen}, R. 2017, \apj, 834, 40

\bibitem[{{Johnson} {et~al.}(2017{\natexlab{a}}){Johnson}, {Caldwell}, {Rich},
  {Mateo}, {Bailey}, {Clarkson}, {Olszewski}, \& {Walker}}]{johnson17-m19}
{Johnson}, C.~I., {Caldwell}, N., {Rich}, R.~M., {et~al.} 2017{\natexlab{a}},
  \apj, 836, 168

\bibitem[{{Johnson} {et~al.}(2017{\natexlab{b}}){Johnson}, {Caldwell}, {Rich},
  {Mateo}, {Bailey}, {Olszewski}, \& {Walker}}]{johnson17-ngc5986}
---. 2017{\natexlab{b}}, \apj, 842, 24

\bibitem[{{Johnson} {et~al.}(2017{\natexlab{c}}){Johnson}, {Caldwell}, {Rich},
  \& {Walker}}]{johnson17-ngc6229}
{Johnson}, C.~I., {Caldwell}, N., {Rich}, R.~M., \& {Walker}, M.~G.
  2017{\natexlab{c}}, \aj, 154, 155

\bibitem[{{Johnson} \& {Pilachowski}(2010)}]{johnson10}
{Johnson}, C.~I., \& {Pilachowski}, C.~A. 2010, \apj, 722, 1373

\bibitem[{{Kacharov} {et~al.}(2013){Kacharov}, {Koch}, \&
  {McWilliam}}]{kacharov13}
{Kacharov}, N., {Koch}, A., \& {McWilliam}, A. 2013, \aap, 554, A81

\bibitem[{Kirby {et~al.}(2008)Kirby, Guhathakurta, \& Sneden}]{kirby08}
Kirby, E.~N., Guhathakurta, P., \& Sneden, C. 2008, The Astrophysical Journal,
  682, 1217

\bibitem[{{Kobayashi} {et~al.}(2011){Kobayashi}, {Karakas}, \&
  {Umeda}}]{kobayashi11}
{Kobayashi}, C., {Karakas}, A.~I., \& {Umeda}, H. 2011, \mnras, 414, 3231

\bibitem[{{Kobayashi} {et~al.}(2006){Kobayashi}, {Umeda}, {Nomoto}, {Tominaga},
  \& {Ohkubo}}]{kobayashi06}
{Kobayashi}, C., {Umeda}, H., {Nomoto}, K., {Tominaga}, N., \& {Ohkubo}, T.
  2006, \apj, 653, 1145

\bibitem[{{Krumholz} {et~al.}(2006){Krumholz}, {Matzner}, \&
  {McKee}}]{krumholz06}
{Krumholz}, M.~R., {Matzner}, C.~D., \& {McKee}, C.~F. 2006, \apj, 653, 361

\bibitem[{{Krumholz} \& {Tan}(2007)}]{krumholztan07}
{Krumholz}, M.~R., \& {Tan}, J.~C. 2007, \apj, 654, 304

\bibitem[{{Lada} {et~al.}(2012){Lada}, {Forbrich}, {Lombardi}, \&
  {Alves}}]{lada12}
{Lada}, C.~J., {Forbrich}, J., {Lombardi}, M., \& {Alves}, J.~F. 2012, \apj,
  745, 190

\bibitem[{{Lada} \& {Lada}(2003)}]{lada03}
{Lada}, C.~J., \& {Lada}, E.~A. 2003, \araa, 41, 57

\bibitem[{{Larson}(1981)}]{larson81}
{Larson}, R.~B. 1981, \mnras, 194, 809

\bibitem[{{Liu} {et~al.}(2017){Liu}, {Ruchti}, {Feltzing}, \& {Primas}}]{liu17}
{Liu}, C., {Ruchti}, G., {Feltzing}, S., \& {Primas}, F. 2017, \aap, 601, A31

\bibitem[{{Marino} {et~al.}(2009){Marino}, {Milone}, {Piotto}, {Villanova},
  {Bedin}, {Bellini}, \& {Renzini}}]{marino09}
{Marino}, A.~F., {Milone}, A.~P., {Piotto}, G., {et~al.} 2009, \aap, 505, 1099

\bibitem[{{Marino} {et~al.}(2015){Marino}, {Milone}, {Karakas}, {Casagrande},
  {Yong}, {Shingles}, {Da Costa}, {Norris}, {Stetson}, {Lind}, {Asplund},
  {Collet}, {Jerjen}, {Sbordone}, {Aparicio}, \& {Cassisi}}]{marino15}
{Marino}, A.~F., {Milone}, A.~P., {Karakas}, A.~I., {et~al.} 2015, \mnras, 450,
  815

\bibitem[{{Marks} {et~al.}(2008){Marks}, {Kroupa}, \& {Baumgardt}}]{marks08}
{Marks}, M., {Kroupa}, P., \& {Baumgardt}, H. 2008, \mnras, 386, 2047

\bibitem[{{Martin} {et~al.}(2002){Martin}, {Kobulnicky}, \&
  {Heckman}}]{martin02}
{Martin}, C.~L., {Kobulnicky}, H.~A., \& {Heckman}, T.~M. 2002, \apj, 574, 663

\bibitem[{{Massari} {et~al.}(2014){Massari}, {Mucciarelli}, {Ferraro},
  {Origlia}, {Rich}, {Lanzoni}, {Dalessandro}, {Valenti}, {Ibata}, {Lovisi},
  {Bellazzini}, \& {Reitzel}}]{massari14}
{Massari}, D., {Mucciarelli}, A., {Ferraro}, F.~R., {et~al.} 2014, \apj, 795,
  22

\bibitem[{{Matzner}(2002)}]{matzner02}
{Matzner}, C.~D. 2002, \apj, 566, 302

\bibitem[{{Mieske} {et~al.}(2006){Mieske}, {Jord{\'a}n}, {C{\^o}t{\'e}},
  {Kissler-Patig}, {Peng}, {Ferrarese}, {Blakeslee}, {Mei}, {Merritt}, {Tonry},
  \& {West}}]{mieske06}
{Mieske}, S., {Jord{\'a}n}, A., {C{\^o}t{\'e}}, P., {et~al.} 2006, \apj, 653,
  193

\bibitem[{{Mieske} {et~al.}(2010){Mieske}, {Jord{\'a}n}, {C{\^o}t{\'e}},
  {Peng}, {Ferrarese}, {Blakeslee}, {Mei}, {Baumgardt}, {Tonry}, {Infante}, \&
  {West}}]{mieske10}
---. 2010, \apj, 710, 1672

\bibitem[{{Milone} {et~al.}(2017){Milone}, {Piotto}, {Renzini}, {Marino},
  {Bedin}, {Vesperini}, {D'Antona}, {Nardiello}, {Anderson}, {King}, {Yong},
  {Bellini}, {Aparicio}, {Barbuy}, {Brown}, {Cassisi}, {Ortolani}, {Salaris},
  {Sarajedini}, \& {van der Marel}}]{milone17}
{Milone}, A.~P., {Piotto}, G., {Renzini}, A., {et~al.} 2017, \mnras, 464, 3636

\bibitem[{{Motte} {et~al.}(2018){Motte}, {Nony}, {Louvet}, {Marsh}, {Bontemps},
  {Whitworth}, {Men'shchikov}, {Nguyen Luong}, {Csengeri}, {Maury}, {Gusdorf},
  {Chapillon}, {K{\"o}nyves}, {Schilke}, {Duarte-Cabral}, {Didelon}, \&
  {Gaudel}}]{motte18}
{Motte}, F., {Nony}, T., {Louvet}, F., {et~al.} 2018, Nature Astronomy, 2, 478

\bibitem[{{Mucciarelli} {et~al.}(2016){Mucciarelli}, {Dalessandro}, {Massari},
  {Bellazzini}, {Ferraro}, {Lanzoni}, {Lardo}, {Salaris}, \&
  {Cassisi}}]{mucciarelli16}
{Mucciarelli}, A., {Dalessandro}, E., {Massari}, D., {et~al.} 2016, \apj, 824,
  73

\bibitem[{{Mura-Guzm{\'a}n} {et~al.}(2018){Mura-Guzm{\'a}n}, {Villanova},
  {Mu{\~n}oz}, \& {Tang}}]{muraguzman18}
{Mura-Guzm{\'a}n}, A., {Villanova}, S., {Mu{\~n}oz}, C., \& {Tang}, B. 2018,
  \mnras, 474, 4541

\bibitem[{{Murray}(2009)}]{murray09}
{Murray}, N. 2009, \apj, 691, 946

\bibitem[{{Murray}(2011)}]{murray11}
---. 2011, \apj, 729, 133

\bibitem[{{Murray} {et~al.}(2010){Murray}, {Quataert}, \&
  {Thompson}}]{murray10}
{Murray}, N., {Quataert}, E., \& {Thompson}, T.~A. 2010, \apj, 709, 191

\bibitem[{{Nomoto} {et~al.}(2013){Nomoto}, {Kobayashi}, \&
  {Tominaga}}]{nomoto13}
{Nomoto}, K., {Kobayashi}, C., \& {Tominaga}, N. 2013, Annual Review of
  Astronomy and Astrophysics, 51, 457

\bibitem[{{Nomoto} {et~al.}(2006){Nomoto}, {Tominaga}, {Umeda}, {Kobayashi}, \&
  {Maeda}}]{nomoto06}
{Nomoto}, K., {Tominaga}, N., {Umeda}, H., {Kobayashi}, C., \& {Maeda}, K.
  2006, \nphysa, 777, 424

\bibitem[{{Paxton} {et~al.}(2011){Paxton}, {Bildsten}, {Dotter}, {Herwig},
  {Lesaffre}, \& {Timmes}}]{MESA11}
{Paxton}, B., {Bildsten}, L., {Dotter}, A., {et~al.} 2011, \apjs, 192, 3

\bibitem[{{Pecaut} {et~al.}(2012){Pecaut}, {Mamajek}, \& {Bubar}}]{pecaut12}
{Pecaut}, M.~J., {Mamajek}, E.~E., \& {Bubar}, E.~J. 2012, \apj, 746, 154

\bibitem[{{Pignatari} {et~al.}(2016){Pignatari}, {Herwig}, {Hirschi},
  {Bennett}, {Rockefeller}, {Fryer}, {Timmes}, {Ritter}, {Heger}, {Jones},
  {Battino}, {Dotter}, {Trappitsch}, {Diehl}, {Frischknecht}, {Hungerford},
  {Magkotsios}, {Travaglio}, \& {Young}}]{nugrid16}
{Pignatari}, M., {Herwig}, F., {Hirschi}, R., {et~al.} 2016, \apjs, 225, 24

\bibitem[{{Piotto} {et~al.}(2015){Piotto}, {Milone}, {Bedin}, {Anderson},
  {King}, {Marino}, {Nardiello}, {Aparicio}, {Barbuy}, {Bellini}, {Brown},
  {Cassisi}, {Cool}, {Cunial}, {Dalessandro}, {D'Antona}, {Ferraro}, {Hidalgo},
  {Lanzoni}, {Monelli}, {Ortolani}, {Renzini}, {Salaris}, {Sarajedini}, {van
  der Marel}, {Vesperini}, \& {Zoccali}}]{piotto15}
{Piotto}, G., {Milone}, A.~P., {Bedin}, L.~R., {et~al.} 2015, \aj, 149, 91

\bibitem[{{Prantzos}(2008)}]{prantzos08}
{Prantzos}, N. 2008, \aap, 489, 525

\bibitem[{{Renzini}(2013)}]{renzini13}
{Renzini}, A. 2013, \memsai, 84, 162

\bibitem[{{Robles-Valdez} {et~al.}(2017){Robles-Valdez},
  {Rodr{\'{\i}}guez-Gonz{\'a}lez}, {Hern{\'a}ndez-Mart{\'{\i}}nez}, \&
  {Esquivel}}]{roblesvaldez17}
{Robles-Valdez}, F., {Rodr{\'{\i}}guez-Gonz{\'a}lez}, A.,
  {Hern{\'a}ndez-Mart{\'{\i}}nez}, L., \& {Esquivel}, A. 2017, \apj, 835, 136

\bibitem[{{Roccatagliata} {et~al.}(2013){Roccatagliata}, {Preibisch}, {Ratzka},
  \& {Gaczkowski}}]{roccatagliata13}
{Roccatagliata}, V., {Preibisch}, T., {Ratzka}, T., \& {Gaczkowski}, B. 2013,
  \aap, 554, A6

\bibitem[{{Salpeter}(1955)}]{salpeter55}
{Salpeter}, E.~E. 1955, \apj, 121, 161

\bibitem[{{Smith} \& {Brooks}(2008)}]{smithbrooks08}
{Smith}, N., \& {Brooks}, K.~J. 2008, in Handbook of Star Forming Regions,
  Volume II, ed. B.~{Reipurth}, 138

\bibitem[{{Stinson} {et~al.}(2013){Stinson}, {Brook}, {Macci{\`o}}, {Wadsley},
  {Quinn}, \& {Couchman}}]{stinson13}
{Stinson}, G.~S., {Brook}, C., {Macci{\`o}}, A.~V., {et~al.} 2013, \mnras, 428,
  129

\bibitem[{{Tinsley}(1980)}]{tinsley80}
{Tinsley}, B.~M. 1980, \fcp, 5, 287

\bibitem[{{Tremonti} {et~al.}(2004){Tremonti}, {Heckman}, {Kauffmann},
  {Brinchmann}, {Charlot}, {White}, {Seibert}, {Peng}, {Schlegel}, {Uomoto},
  {Fukugita}, \& {Brinkmann}}]{tremonti04}
{Tremonti}, C.~A., {Heckman}, T.~M., {Kauffmann}, G., {et~al.} 2004, \apj, 613,
  898

\bibitem[{{Vanzi} \& {Sauvage}(2004)}]{vanzi04}
{Vanzi}, L., \& {Sauvage}, M. 2004, \aap, 415, 509

\bibitem[{{Villanova} {et~al.}(2017){Villanova}, {Moni Bidin}, {Mauro},
  {Munoz}, \& {Monaco}}]{villanova17}
{Villanova}, S., {Moni Bidin}, C., {Mauro}, F., {Munoz}, C., \& {Monaco}, L.
  2017, \mnras, 464, 2730

\bibitem[{{Willman} \& {Strader}(2012)}]{willmanstrader12}
{Willman}, B., \& {Strader}, J. 2012, \aj, 144, 76

\bibitem[{{Zhang} {et~al.}(2018){Zhang}, {Romano}, {Ivison}, {Papadopoulos}, \&
  {Matteucci}}]{zhang18}
{Zhang}, Z.-Y., {Romano}, D., {Ivison}, R.~J., {Papadopoulos}, P.~P., \&
  {Matteucci}, F. 2018, \nat, 558, 260

\end{thebibliography}

\appendix

\section{Stellar Lifetimes}
\label{sec:stellar-lifetimes}
In order to parameterize the stellar mass-lifetime relation, I have used data from the MESA Isochrones and Stellar Tracks (MIST) project \citep{choi16} based on the Modules for Experiments in Stellar Astrophysics (MESA) stellar evolution code \citep{MESA11}. Metal-poor models ([Fe/H]=$-2.0$, $-1.75$, $-1.50$) were downloaded\footnote{http://waps.cfa.harvard.edu/MIST/model\_grids.html}; at the high masses that are relevant, the differences are negligible so the [Fe/H]=$-1.75$ models were used. The adopted ages correspond to the end of Carbon burning; later evolutionary phases are assumed to be short in comparison.

\begin{figure}
\plotone{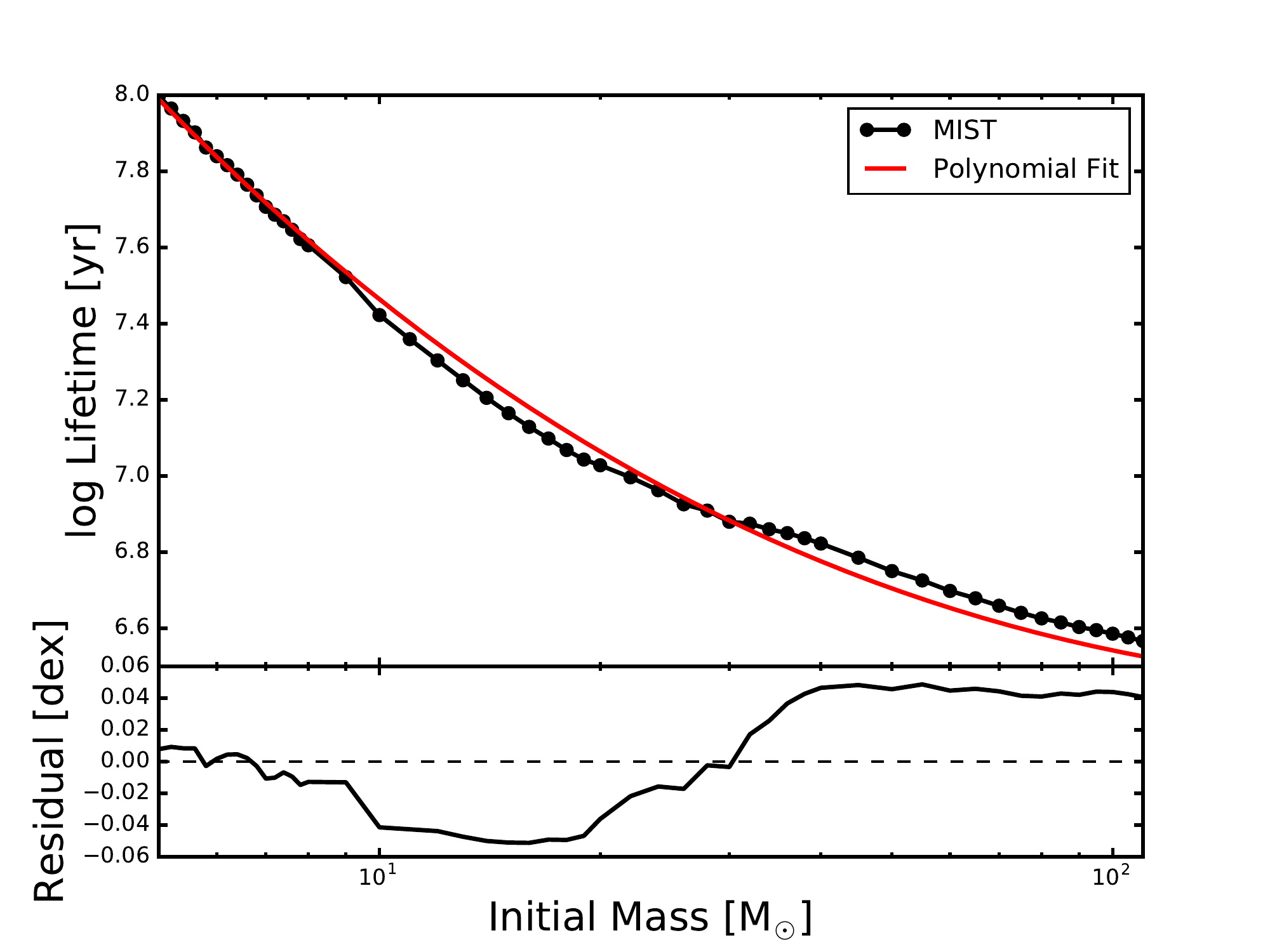}
\caption{\label{fig:stellar-lifetimes}%
Lifetimes of massive metal-poor ([Fe/H]=$-1.75$) stars from MIST \citep{choi16}. The polynomial fit of equation~(\ref{eq:lifetime-polyfit}) is good to 0.05~dex (12\%) for all masses with lifetimes less than the age of the Universe.}
\end{figure}

The stellar mass-lifetime relation is shown in Figure~\ref{fig:stellar-lifetimes}. A polynomial approximation has been fit to the high mass ($m > 5 M_{\odot}$) data:
\begin{equation}\label{eq:lifetime-polyfit}
    \log \tau = -0.086 \left(\log m\right)^3 + 0.095 \left(\log m\right)^2 - 3.17 \left(\log m\right) + 9.77
\end{equation}
for initial stellar mass $m$ and lifetime $\tau$. This fit is good to 0.05~dex (12\%) for all masses $m>0.6 M_{\odot}$, at which point the lifetime is significantly longer than the age of the Universe. Equation~(\ref{eq:lifetime-polyfit}) is the adopted stellar mass-lifetime relation.

\section{Core Collapse Supernova Yields}\label{sec:yields}
Heavy element yields from CCSNe have large theoretical uncertainties due to the difficulty of performing multidimensional high resolution hydrodynamic simulations with neutrino transport and nucleosynthesis with large reaction networks. I have tested yields from four different sources:
\begin{enumerate}
 \item The \citetalias{bailin09} parametrization
\begin{equation}\label{eq:bh09-yield-fit}
    m_Z(m) = (B + C m) m
\end{equation}
with $B = 1.18\%$, $C = 0.548\%$, and all masses in units of \Msun.

 \item The yield tables for core collapse supernovae from \citet{nomoto13}\footnote{http://star.herts.ac.uk/$\sim$chiaki/works/YIELD\_CK13.DAT}, which collates models produced by this group \citep{nomoto06,kobayashi06,kobayashi11}. These cover progenitor masses from 10 -- 40~$M_{\odot}$ and a range of metallicities.

 \item NuGrid \citep{nugrid16} CCSN simulations that cover a range of progenitor masses and metallicities\footnote{http://www.nugridstars.org/data-and-software/yields.}.
 
 \item \citet{cote16} CCSN simulations that cover a range of progenitor masses and metallicities\footnote{Data kindly provided by C.~West and A.~Heger.}. 
\end{enumerate}

\begin{figure*}
\plotone{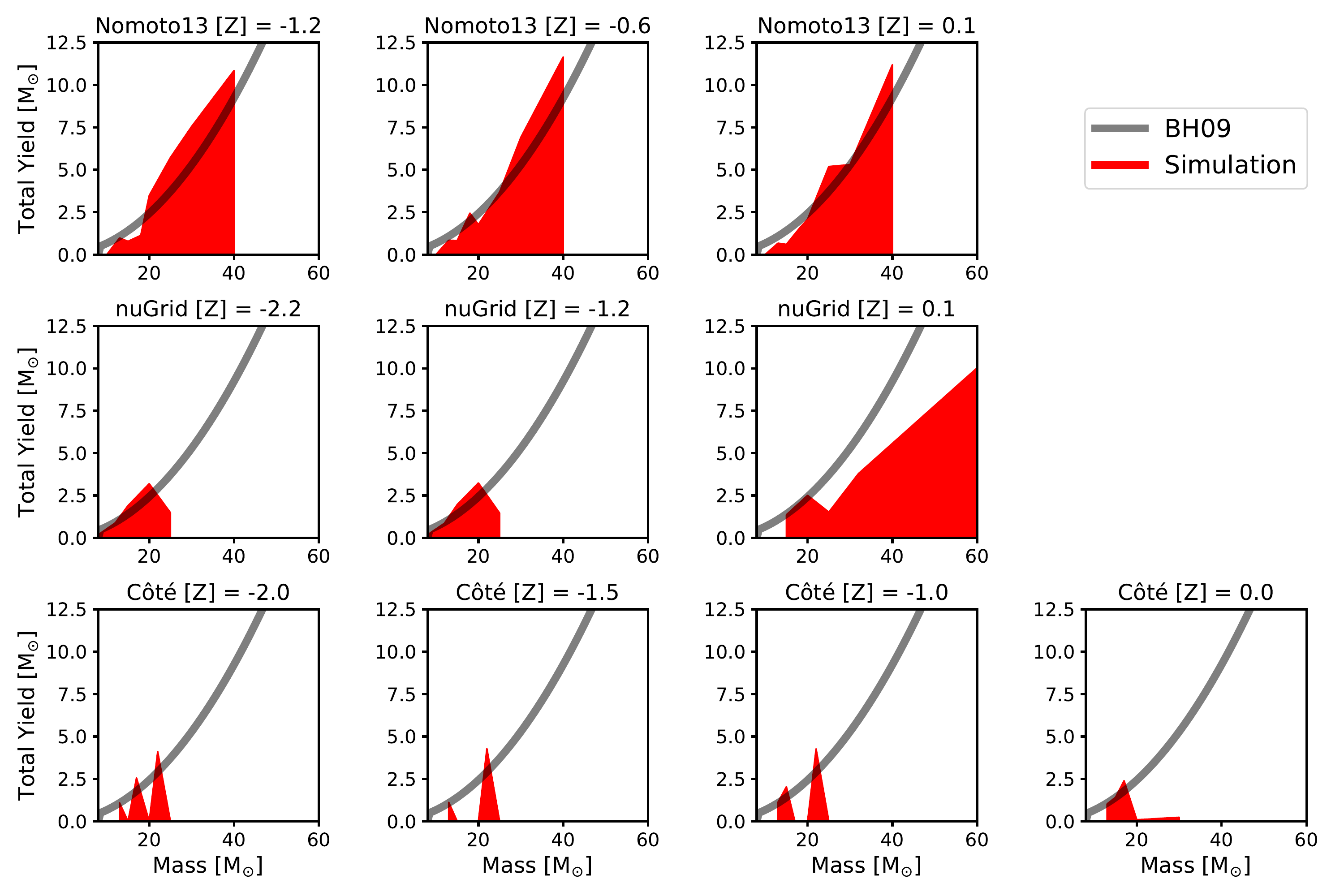}
\caption{\label{fig:yield-comparison}%
Total metal yield, in solar masses, for a variety of progenitor masses and metallicities from \citet{nomoto13}, nuGrid \citep{nugrid16} and \citet{cote16}. Metallicities are expressed as $[Z] \equiv \log Z/Z_{\odot}$. The parameterization from \citetalias{bailin09} is shown for comparison. Vertical edges indicate the edge of the model grid in mass.}
\end{figure*}

The total metal yield for relevant models are shown in Figure~\ref{fig:yield-comparison}. A few trends are apparent:
\begin{itemize}
 \item A large fraction of the \citet{cote16} models collapse straight to a black hole and yield \textbf{no} heavy elements. The details of which models explode show no simple pattern, but the relatively common GC metallicity [Z]=$-1.5$ is particularly prone to collapsing directly to a black hole.
 The models that do explode follow the \citetalias{bailin09} fit quite well below $25 M_{\odot}$ at most metallicities.
 
 \item The \citet{nomoto13} and nuGrid models all explode by construction. They also follow the \citetalias{bailin09} fit up to $20 M_{\odot}$, but the nuGrid models have lower yields for higher mass progenitors due to fallback, while the \citet{nomoto13} models match the \citetalias{bailin09} fit well at most masses and metallicities.
 
 \item The differences between simulation groups completely overwhelms any metallicity differences within any one group's models, so we can neglect metallicity effects as a significant contributor to the uncertainty
\end{itemize}

\begin{figure*}
\plottwo{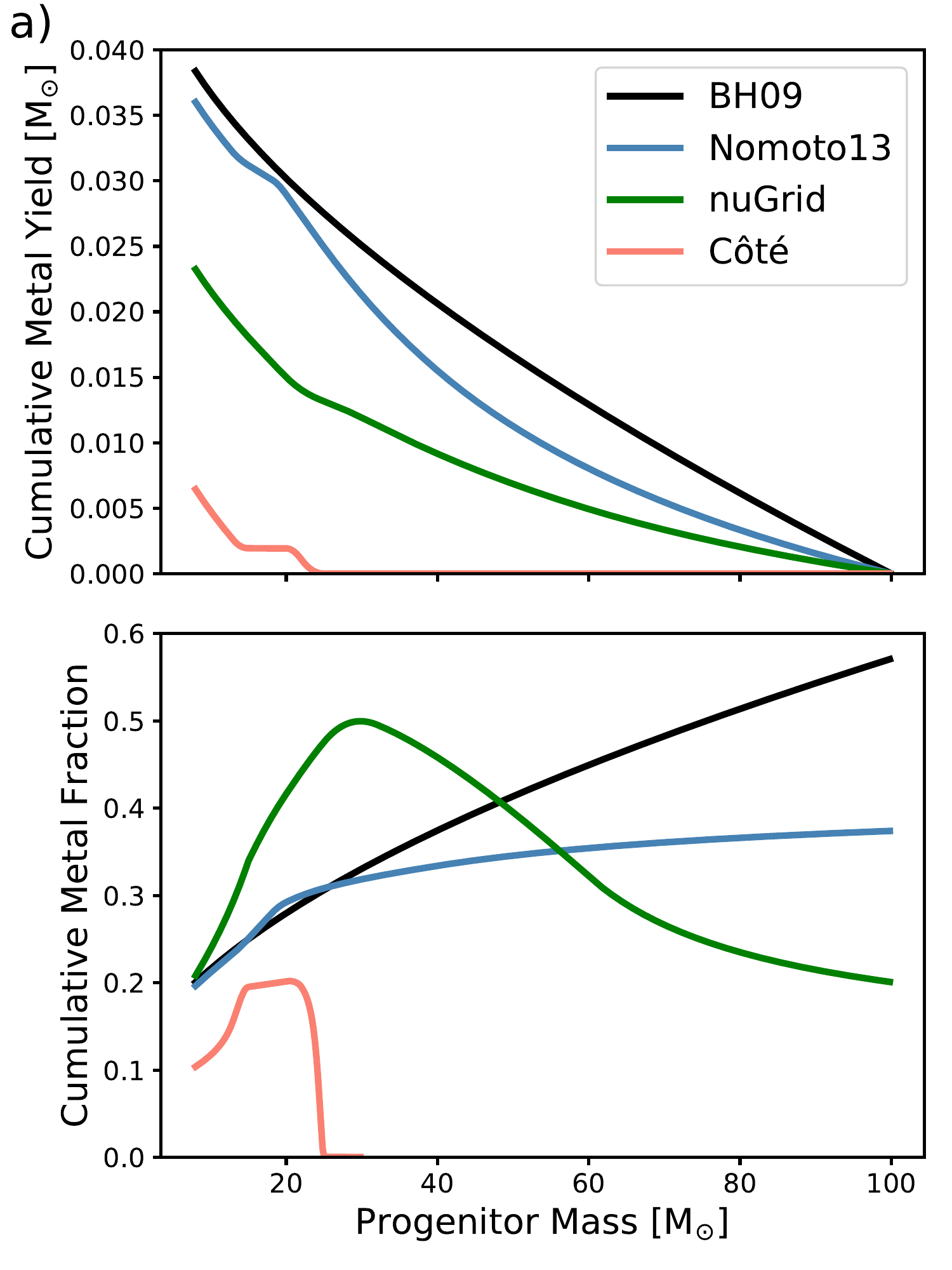}{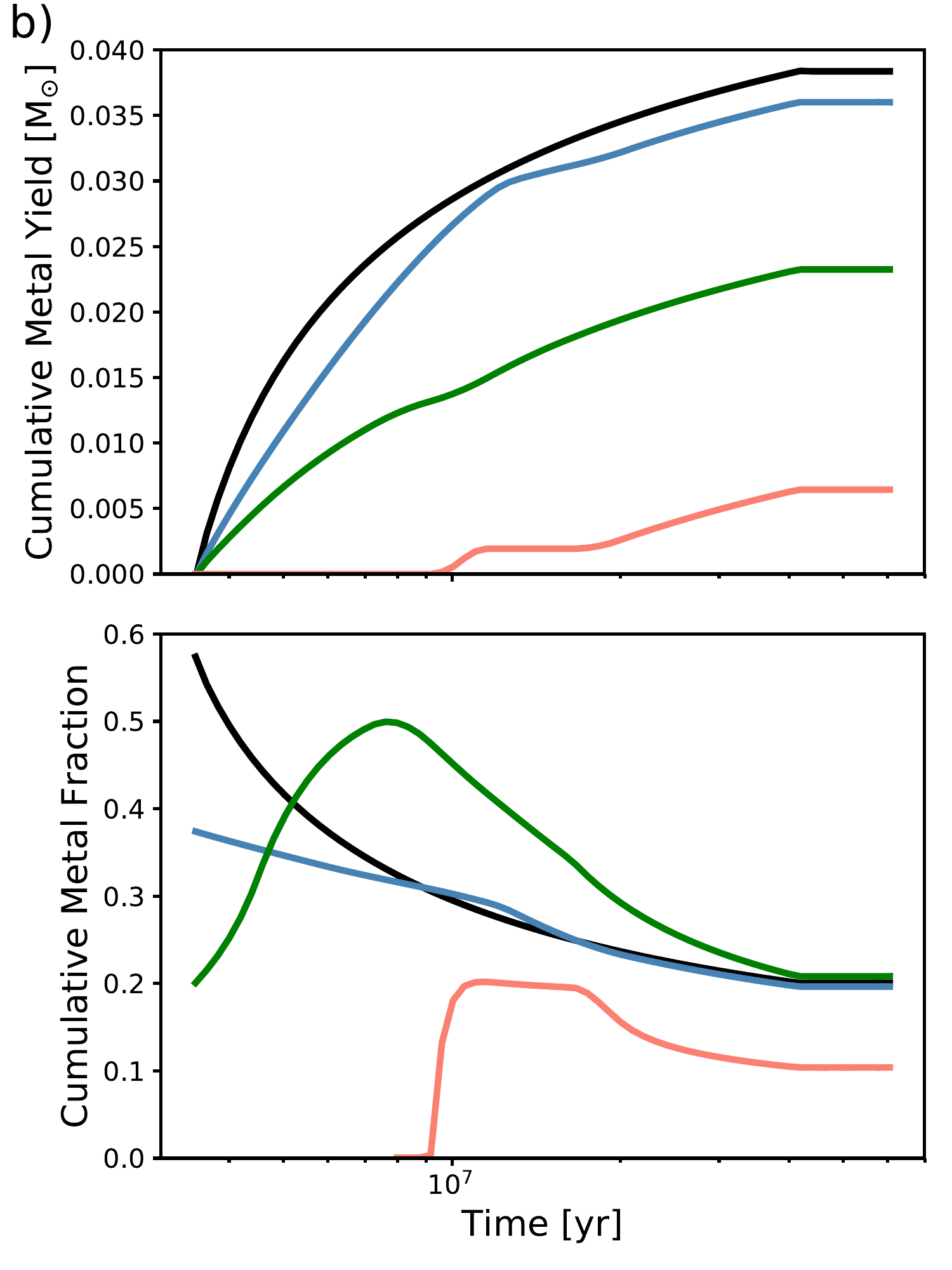}
\caption{\label{fig:yield-functions}%
\textit{a)} Cumulative metal yield (top) from stars at least as massive as the plotted progenitor mass, per solar mass of stars formed with the truncated Salpeter IMF, for the different CCSN yield models. The bottom panel shows the net metallicity of all ejecta from stars at least as massive as the plotted progenitor mass.
\textit{b)} Cumulative metal yield (top) produced as a function of time, per solar mass of stars formed with the truncated Salpeter IMF. The bottom panel shows the net metallicity of all ejecta produced as a function of time.%
}
\end{figure*}

Given the large uncertainty in the yields, and the significant effect that this could have on the GC self enrichment model, I test four different models for CCSN yields:
\begin{description}
 \item[BH09] The \citetalias{bailin09} parameterization (Equation~\ref{eq:bh09-yield-fit}) for the metal mass. The total (H + He + metals) ejected mass assumes that all supernovae leave $2~M_{\odot}$ progenitors; given that this is much less than the progenitor mass, the exact value does not make a significant difference.
 \item[Nomoto13] The $[Z]=-1.2$ \citet{nomoto13} model, extrapolated at $8 < M < 10~M_{\odot}$ using the BH09 model and at $M > 40 M_{\odot}$ by assuming that the same fraction of the initial mass is released, and is released as metals, as in the $40 M_{\odot}$ model.
 \item[nuGrid] The solar metallicity nuGrid model, extrapolated at $8 < M <15~M_{\odot}$ using the BH09 model and at $M > 60 M_{\odot}$ by assuming that the same fraction of the initial mass is released, and is released as metals, as in the $60 M_{\odot}$ model.
 I have adopted the solar metallicity models because (a) they extend to higher masses, which is important since nearly half of the total metal yield comes from models more massive than the most massive ($25 M_{\odot}$) lower-metallicity model, and (b) although the relative yield of different elements can change significantly for different metallicities, the total metal yield is quite insensitive to metallicity for overlapping parts of the parameter grid.
 \item[C\^ot\'e] The $[Z]=-1.5$ \citet{cote16} model, extrapolated at $8 < M < 13~M_{\odot}$ using the BH09 model and at $M > 25 M_{\odot}$ by assuming they all collapse directly to a black hole, i.e. zero yield.
\end{description}
These adopted yield functions are plotted in Figure~\ref{fig:yield-functions}, for a stellar population described by the truncated Salpeter IMF and the stellar lifetimes derived in Appendix~\ref{sec:stellar-lifetimes}. The BH09, Nomoto13, and nuGrid models all produce heavy elements over a similar timescale but with different normalizations (the BH09 model produces nearly twice as much metal as the nuGrid model, with Nomoto13 lying in between).
The C\^ot\'e model, on the other hand, not only produces much less metal mass, but does it from much lower mass stars and therefore at much later times, with none produced within the first 10~Myr.
The net metallicity of the ejecta varies by a factor of $\sim 2$.

\end{document}